\documentclass[a4paper,fleqn,usenatbib,useAMS]{mnras}

\usepackage{graphicx}	
\usepackage{amsmath}	
\usepackage{amssymb}	
\usepackage{multicol}        
\usepackage{bm}	
\usepackage{pdflscape}	
\usepackage{subfigure}
\usepackage{longtable}
\usepackage{array}
\usepackage{booktabs}

\usepackage[T1]{fontenc}
\usepackage{ae,aecompl}

\title[Jet Propagation in Expanding Medium for GRBs]{Jet Propagation in Expanding Medium for Gamma-Ray Bursts}

\author[Hamidani \& Ioka]{Hamid Hamidani$^1$\thanks{E-mail: hamidani.hamid@yukawa.kyoto-u.ac.jp} and Kunihito Ioka$^1$
\\
$^{1}$ Yukawa Institute for Theoretical Physics, Kyoto University, Kyoto 606-8502, Japan\\
}

\date{Accepted XXX. Received YYY; in original form ZZZ}

\pubyear{2020}

\begin{document}
\label{firstpage}
\pagerange{\pageref{firstpage}--\pageref{lastpage}}
\maketitle

\begin{abstract}
The binary neutron star (BNS) merger event GW170817 clearly shows that a BNS merger launches a short Gamma-Ray Burst (\textit{s}GRB) jet.
Unlike collapsars, where the ambient medium is static, in BNS mergers the jet propagates through the merger ejecta that is expanding outward at substantial velocities ($\sim 0.2c$).
Here, we present semi-analytic and analytic models to solve the propagation of GRB jets through their surrounding media.
These models improve our previous model by including the jet collimation by the cocoon self-consistently.
We also perform a series of 2D numerical simulations of jet propagation in BNS mergers and in collapsars to test our models. 
Our models are consistent with numerical simulations in every aspect (the jet head radius, the cocoon's lateral width, the jet opening angle including collimation, the cocoon pressure, and the jet-cocoon morphology).
The energy composition of the cocoon is found to be different depending on whether the ambient medium is expanding or not; in the case of BNS merger jets, the cocoon energy is dominated by kinetic energy, while it is dominated by internal energy in collapsars.
Our model will be useful for estimating electromagnetic counterparts to gravitational waves. 
\end{abstract}



\section{Introduction}
Observation of the gravitational wave signal from the binary neutron star (BNS) merger event GW170817 by the Laser Interferometer Gravitational-Wave Observatory (LIGO) and the Virgo Consortium (LVC) [\citealt{2017PhRvL.119p1101A}], and the follow-up observation campaign across the electromagnetic (EM) spectrum marked the dawn of the era of multi-messenger astronomy (\citealt{2017ApJ...848L..13A}).
One of the most important findings was the association of GW170817 with the prompt emission of the short Gamma-Ray Burst (\textit{s}GRB) \textit{s}GRB 170817A $\sim 1.7$ s after the GW signal (\citealt{2017ApJ...848L..13A}).
Clear evidences of a relativistic jet have also been obtained from radio observations at later times (\citealt{2018Natur.561..355M}).
As GRB observations show, relativistic jets are very important in time-domain astronomy, especially because of the EM emission over a wide spectrum (i.e., the prompt and the afterglow emission).

This scenario, that the BNS merger powers a relativistic jet, was theoretically suggested for \textit{s}GRBs in the past (\citealt{1986ApJ...308L..43P}; \citealt{1986ApJ...308L..47G}; \citealt{1989Natur.340..126E}).
Recent studies of numerical relativity show that after the merger, 
a central engine is formed and is surrounded by $\sim 10^{-3} - 10^{-2} M_\odot$ of matter that has been ejected during the merger, referred to as ``the dynamical ejecta'', that expands at a speed of $\sim 0.2c$, where $c$ is the light speed (\citealt{1999PhRvD..60j4052S}; \citealt{2000PhRvD..61f4001S}; \citealt{2013PhRvD..87b4001H}; \citealt{2013ApJ...773...78B}; \citealt{2018ApJ...869..130R}; etc.).
Also, later on ($< 10$ s after the merger), matter gets ejected from the torus surrounding the merger remnant in the form of wind  (\citealt{2017PhRvL.119w1102S,2018ApJ...869L..35R,2018ApJ...860...64F,2020arXiv200700474F,2020arXiv200804333N,2020arXiv200411298C}; etc.).

Multi-messenger observations of the BNS merger event GW170817 indicates that 
the relativistic jet was launched from the central engine within $\sim 1.7$ s after the merger (more precisely, within $\sim 1.3$ s after the merger according to \citealt{2020MNRAS.491.3192H}, see their Figure 9; within $0.4$ s according to \citealt{2020arXiv200410210L}).
Therefore, the jet must have propagated through the dense surrounding medium (i.e., the dynamical ejecta), and successfully broke out of it for the \textit{s}GRB 170817A to be emitted (as it has been observed).
This is because the jet outflow is not observable unless it propagates up to the outer edge of the medium and eventually breaks out of it;
as it is the case for long GRBs, where the relativistic jet propagates through the stellar envelope of a massive star.

During its propagation,
the jet continuously injects energy into the expanding ejecta material.
This produces the hot cocoon. 
The cocoon immediately surrounds the jet, interacts with it, and collimates it.
Although this phase, where the jet is confined inside the ejecta, is short, it is critical, as it shapes the jet (and the cocoon) structure (see \citealt{2020arXiv200602466G}).
After the breakout, the jet (and the cocoon) is the source of different EM counterparts over a wide band, 
and it is the key to interpreting them (\citealt{2017ApJ...834...28N,2017ApJ...848L...6L,2018MNRAS.473..576G,2018PhRvL.120x1103L,2018ApJ...855..103P,2018ApJ...867...18N,2018PTEP.2018d3E02I}; etc.).
Also, this connection between the jet and the EM counterparts allows us to make use of observational data to extract crucial information and better understand the phenomenon of \textit{s}GRBs (e.g., the jet angular structure, \citealt{2019MNRAS.489.1919T,2020MNRAS.tmp.2102T}; 
the property of the jet outflow, \citealt{2019MNRAS.487.4884I}; the central engine, \citealt{2019ApJ...876..139G,2020MNRAS.491.3192H,2020arXiv200410210L,2020arXiv200607376S}; the physics of neutron density matter, \citealt{2019ApJ...881...89L}; the viewing angle, \citealt{2020arXiv200501754N}; etc.).
Therefore, 
the jet propagation through the ejecta surrounding the BNS merger remnant, until the breakout, is a key process in \textit{s}GRBs (as it is in collapsars and long GRBs).

The propagation of astrophysical jets through dense ambient media has been the subject of intensive theoretical works; mostly, in the context of Active Galactic Nuclei (AGNs) and collapsars (\citealt{1989ApJ...345L..21B}; \citealt{1997ApJ...479..151M}; \citealt{1999ApJ...524..262M}; \citealt{2003MNRAS.345..575M}; \citealt{2011ApJ...740..100B}; \citealt{2013ApJ...777..162M}; \citealt{2018MNRAS.477.2128H}; etc.).
One critical difference in the context of BNS mergers is that the ambient medium expands at substantial velocities (\citealt{2013PhRvD..87b4001H}), while it is static in AGNs and collapsars (\citealt{1999ApJ...524..262M}).
This further complicates the problem of modeling the jet propagation in BNS mergers.

There have been an increasing number of studies dedicated to solving jet propagation in BNS mergers through numerical simulations, especially after the discovery of GW170817 (\citealt{2014ApJ...784L..28N}; \citealt{2014ApJ...788L...8M}; \citealt{2015ApJ...813...64D}; \citealt{2018MNRAS.475.2971B}; \citealt{2018ApJ...866....3D}; \citealt{2017ApJ...848L...6L}; \citealt{2018MNRAS.473..576G}; \citealt{2018MNRAS.479..588G}; \citealt{2018ApJ...863...58X}; \citealt{2020MNRAS.495.3780N}; \citealt{2020arXiv200602466G}; etc.).
However, the subject is still far from being well understood.

Using ideas from the modeling of the jet-cocoon in collapsars (e.g., \citealt{2011ApJ...740..100B}; \citealt{2013ApJ...777..162M}; \citealt{2018MNRAS.477.2128H}), several studies presented analytic modeling of the jet-cocoon in an expanding medium (\citealt{2017MNRAS.471.1652L}; \citealt{2018MNRAS.475.2659M}; \citealt{2018ApJ...866....3D}; \citealt{2018ApJ...866L..16M}; \citealt{2019ApJ...876..139G}; \citealt{2020A&A...636A.105S}; \citealt{2020MNRAS.491.3192H}; \citealt{2020MNRAS.491..483L}; \citealt{2020ApJ...895L..33B}; etc.).
Although some of these works offer promising results, many of them overlooked important aspects, such as the jet collimation, the expansion of the ejecta and its effect on the cocoon (e.g., on the cocoon pressure and on the cocoon radius), etc. 
And still, there is no analytic model for the jet propagation in an expanding medium that is simple to use, and robust at the same time (which is necessary to investigate other related topics, 
such as the emission from the cocoon).

This work presents analytic modeling of jet propagation in an expanding medium.
This model is an upgrade of the model presented in \citet{2020MNRAS.491.3192H}.
The main improvement is that the jet collimation (i.e., the jet opening angle) is calculated in a self-consistent manner by using the cocoon pressure; rather than relying on the assumption of a constant opening angle, and on the free parameter $f_j$.
Another addition is the semi-analytic model presented here, where results are found through numerical integration of ordinary differential equations, relying less on approximations.

The aim of our work is to present a physical model that accurately describes the jet-cocoon system in an expanding medium, in consistency with numerical simulation.
A crucial point in our study is that our modeling is based on rigorous analysis of the jet-cocoon system in numerical simulations (in both expanding and static media), which shows that the expansion of the medium does intrinsically affect the jet-cocoon (e.g., the energy composition of the cocoon, the expansion velocity of the cocoon, etc.).
We show that the jet-cocoon system, in an expanding medium, can be described by a set of equations that can be solved numerically (referred to as the ``semi-analytic" solution).
We also show that, with some reasonable approximations, the system of equations can be simplified and solved analytically (the ``analytic" solution).
Both solutions are rigorously compared to the results from the numerical simulations and found consistent.

This paper is organized as follows. 
In Section \ref{sec:2}, physical modeling of the jet-cocoon system in both expanding and static media is presented, and two (semi-analytic and analytic) solutions are derived. 
In Section \ref{sec:3}, numerical simulations are presented and compared to both solutions.
A conclusion is presented in Section \ref{sec:conclusion}.

\section{The jet-cocoon physical model}
\label{sec:2}
The jet-cocoon model presented here is an upgrade of previous models, in particular in  \citet{2011ApJ...740..100B}; \citet{2013ApJ...777..162M}; \citet{2018MNRAS.477.2128H}; \citet{2020MNRAS.491.3192H}; etc.
Unlike previous models, this model allows us to treat the jet collimation by the cocoon in expanding media.
Therefore, this model can be applied not only to the case of collapsar jets (where the jet propagates through the static stellar envelope of a massive star) but also to the case of BNS merger jets (where the jet propagates through the expanding dynamical ejecta).
This model is also an upgrade of the model presented \citet{2020MNRAS.491.3192H}; 
it takes into account the cocoon and its pressure on the jet (i.e., collimation), hence allowing to derive the jet opening angle in a self-consistent manner.

\subsection{Jump conditions}
\label{sec:Jump conditions}
The jet head dynamics in a dense ambient medium can be determined by the shock jump conditions at the jet's head (e.g., \citealt{1989ApJ...345L..21B}; \citealt{1997ApJ...479..151M}; \citealt{2003MNRAS.345..575M}):
\begin{eqnarray}
h_j \rho_j c^2 (\Gamma\beta)_{jh}^2 + P_j = h_a \rho_a c^2 (\Gamma\beta)_{ha}^2 + P_a ,
\label{eq:jump}
\end{eqnarray}
where $h$, $\rho$, $\Gamma=(1-\beta^2)^{-1/2}$, and $P$ are enthalpy, density, Lorentz factor (with $\beta$ being the velocity normalized by the light speed), and pressure of each fluid element, all measured in the fluid's rest frame. 
The subscripts $j$, $h$, and $a$ refer to three domains: the relativistic jet, the jet head, and the cold ambient medium, respectively. Typically, both $P_a$ and $P_j$ in equation (\ref{eq:jump}) are negligible terms;
hence, we can write the jet head velocity as (for more details see \citealt{2018PTEP.2018d3E02I}; \citealt{2020MNRAS.491.3192H}):
\begin{eqnarray}
\beta_h  =  \frac{\beta_j - \beta_a}{1 + \tilde{L}^{-1/2}} +  \beta_a ,
\label{eq:beta_h 1}
\end{eqnarray}
where $\tilde{L}$ is the ratio of energy density between the jet and the ejecta,
$\tilde{L}  =  \frac{h_j \rho_j \Gamma_j^2}{h_a \rho_a \Gamma_a^2}$,
which can be approximated as:
\begin{eqnarray}
\tilde{L} \simeq \frac{L_j}{\Sigma_j(t) \rho_a c^3 \Gamma_a^2}   ,
\label{eq:L expression approx}
\end{eqnarray}
where $L_j$ is the (true) jet luminosity (per one jet), $\Sigma_j(t)=\pi\theta_j^2(t) r_h^2(t)$ is the jet head cross section, with $r_h(t)$ being the jet head radius [$r_h(t)=\int^t_{t_0}\beta_h dt+r_0$], and $\theta_j(t)$ being the jet opening angle (\citealt{2003MNRAS.345..575M}; \citealt{2011ApJ...740..100B}). 

\subsection{Main approximations}
\label{sec:Main approximations}
Here, in our analytic modeling, we consider a similar set of approximations to that in \citet{2020MNRAS.491.3192H}.
In summary: 
\begin{enumerate}
    \item The analytic treatment presented here is limited to the case of a non-relativistic jet head where: 
    \begin{equation}
        \tilde{L}\ll (1-\beta_a)^2.
    \end{equation}
    \item In BNS mergers, during the merger, matter is dynamically ejected from the system (i.e., the dynamical ejecta) and surrounds the later formed central engine.
    After the merger, another plausible source of mass ejection is the wind. 
    The mass of matter driven by the wind could be substantial in early time (e.g., $\sim 0.01 M_\odot$ in the case of magnetically driven wind; see \citealt{2020arXiv200411298C}), however, the launch time (of the wind) may be later than the jet depending on the type of wind (see \citealt{2018ApJ...860...64F} for the case of viscous wind). 
    For simplicity, we consider the case where the central engine is only surrounded by the dynamical ejecta.
    Note that, here, the ambient medium (often referred to as ``the ejecta'') is defined as the medium surrounding the central engine through which the jet propagation takes place; regardless of whether this inner region is gravitationally bound to the central engine or not. 
    Hence, for simplicity, the total mass of the ambient medium, $M_{a}$, is defined by the ambient medium's density through which the jet head propagates $\rho_a(r,t)$ [i.e., in the polar direction]\footnote{Note that, in the case of BNS mergers, the density throughout the dynamical ejecta is angle dependent, with the density in the polar region being much lower than that near the equatorial region (see Figure 8 in \citealt{2020MNRAS.491.3192H}). This effect results in equation (\ref{eq:M_a}) giving a mass, $M_a$, of $\sim 0.002 M_\odot$ if accounting for a total dynamical ejecta mass of $\sim 0.01 M_\odot$ (for more details see \citealt{2020MNRAS.491.3192H}). Note that, this value for the mass $M_a$ can be scaled-up to account for contribution form the wind.}:
    \begin{equation}
        M_{a}=\int_{r_0}^{r_{m}(t)}4\pi r^2 \rho_a(r,t) {\rm d}r,
        \label{eq:M_a}
    \end{equation} 
    where $r_0$ is the inner boundary of the ambient medium, which is of the order of $10^6-10^7$ cm, and $r_m(t)$ is the outer radius of the ambient medium. 
    This definition of $M_a$ is also used for the collapsar case.
    \item In the case of an expanding medium (BNS merger), the approximation of a homologous expansion is used (see Figure 8 in \citealt{2020MNRAS.491.3192H}). Hence, the radial velocity of the ambient medium as a function of radius $r$ and time $t$ is approximated as: 
    \begin{equation}
    v_a(r,t)=\left[\frac{r}{r_m(t)}\right]v_{m},    
    \label{eq:v_a}
    \end{equation}
    with $r_m(t)$ being the outer radius of the ambient medium, and $v_{m}$ being the maximum velocity of the ambient medium at the radius $r_m(t)$. 
    In the case of collapsars, the velocity is negligible, i.e., $v_m=0$. 
    \item The ambient medium's density profile is approximated to a power-law function, with ``$n$" being its index. Hence, considering the homologous expansion of the ambient medium, the density can be written as: 
    \begin{equation}
        \rho_a(r,t)=\rho_0 \left[\frac{r_0}{r}\right]^n \left[\frac{r_{m,0}}{r_m(t)}\right]^{3-n}, 
        \label{eq:rho_a}
    \end{equation} 
    where $\rho_0=\rho_a(r_0,t_0) = \left[\frac{M_a}{4\pi r_0^n}\right]\left[\frac{3-n}{r_{m,0}^{3-n}-r_0^{3-n}}\right]$, and $r_{m,0}=r_m(t_0)$, with $t_0$ being the jet launch time. 
    This expression is simpler in the collapsar case, where $r_m(t)=r_{m,0}$ ($\equiv r_m$).
    Also, the value $n=2$ is assumed for the density profile of the ambient medium, for both the BNS merger case and the collapsar case.
    It should be noted that this is a simplification as, ideally, $n\sim 2-3.5$ in the BNS merger case (see Figure 8 in \citealt{2020MNRAS.491.3192H}), and $n\sim 1.5-3$ in the collapsar case (see Figure 2 in \citealt{2013ApJ...777..162M}).
    Also, it should be noted that the analytic modeling presented here is limited for the case $n<3$ (\citealt{2011ApJ...740..100B}; \citealt{2020MNRAS.491.3192H}).
    \item The pressure in the cocoon, $P_c$, is dominated by radiation pressure. Hence, it can be written as:
    \begin{equation}
        P_c= \frac{E_i}{3 V_c}, 
        \label{eq:P_c}
    \end{equation} 
    where $E_i$ is the cocoon's internal energy and $V_c$ is the cocoon's volume.
    \item Based on rigorous analysis of the cocoon in numerical simulations, we suggest that the cocoon's shape is better approximated to an ellipsoidal (see \citealt{2020MNRAS.491.3192H}; also see Figure \ref{fig:maps} below); where the ellipsoid's semi-major axis and semi-minor axis at a time $t$ are $\frac{1}{2}r_h(t)$ and $r_c(t)$, respectively, with $r_c(t)$ being the cocoon's lateral width (from the jet axis) at the radius $\frac{1}{2}r_h(t)$ [see also equation (\ref{eq:S1})].
    Hence, the volume of the cocoon (in one hemisphere) can be written as:\footnote{Ideally, the jet volume should be subtracted from the above expression of $V_c$ to give a more accurate expression of the cocoon volume. However, as long as the jet opening angle is not very large (as it is the case here), the jet volume can be neglected.}
    \begin{equation}
        V_c=\frac{2\pi}{3}r_c^2(t) r_h(t).
        \label{eq:V_c}
    \end{equation}
    Note that this presents one of the differences compared to previous works -- typically assuming a cylindrical cocoon shape (e.g., \citealt{2011ApJ...740..100B}; \citealt{2013ApJ...777..162M}; \citealt{2020A&A...636A.105S}).
    \item As previously explained in \citet{2013ApJ...777..162M}, and more rigorously in \citet{2018MNRAS.477.2128H}, the analytic description of $\tilde{L}$ in equation (\ref{eq:L expression approx}) needs to be calibrated by numerical simulations. 
    The parameter $N_s$ is introduced to calibrate the analytic value of $\tilde L$ so that:
    \begin{equation}    
        \tilde{L}_c = N_s^2 \tilde{L},
        \label{eq:N_s}
    \end{equation}
where $\tilde{L}_c$ is the calibrated counterpart of $\tilde{L}$.
Here, the value of $N_s$ for the analytic (or semi-analytic) solution is chosen so that the analytic (or semi-analytic) breakout time is calibrated to the breakout time measured in numerical simulations (see Table \ref{Table:sim}).

As previously noted in \citet{2020MNRAS.491.3192H} (also see \citealt{2013ApJ...777..162M,2018MNRAS.477.2128H}), $N_s$ accounts for the part of meandering energy, without contributing to the forward jet head motion.
However, it should be noted that value of $N_s$ should be dependent on the parameter space (in particular on the value of $\tilde{L}$, see Figure 12 in \citealt{2018MNRAS.477.2128H}; and $v_m$, see Appendix \ref{sec:C}).
Therefore, the values of $N_s$ used here should be limited to the parameter space used here, and should not be taken at face value (for more details on $N_s$ refer to Appendix \ref{sec:C}).
\end{enumerate}

Following the introduction of the calibration coefficient $N_s$, $\tilde{L}$ is substituted by $\tilde{L}_c$ in equation (\ref{eq:beta_h 1}), and with $\beta_j\simeq 1$, the jet head velocity can be written as:
\begin{eqnarray}
\beta_h  = \left[ (1 - \beta_a)(1 + \tilde{L}_c^{1/2})^{-1} \right] \tilde{L}_c^{1/2} +  \beta_a ,
\label{eq:beta_h 2}
\end{eqnarray}
where $\tilde{L}_c$ can be found from equations (\ref{eq:L expression approx}) and (\ref{eq:N_s}).
Given the jet luminosity $L_j$, the ambient medium's velocity $\beta_a$ [$=v_a(r,t)/c$; see equation (\ref{eq:v_a})], and the density $\rho_a(r,t)$ [see equation (\ref{eq:rho_a})], the only unknown quantity for $\tilde{L}_c$ (i.e., $\beta_h$) to be determined is the jet head cross-section $\Sigma_j(t)$.
The jet head opening angle $\theta_j(t)$; and hence $\Sigma_j(t)$; will be determined in Section \ref{sec:cocoon collimation} by considering the collimation of the jet by the cocoon.

The jet head velocity [i.e., $\beta_h$ in equation (\ref{eq:beta_h 2})] will be solved in two different ways: Semi-analytically and analytically (details are given in Sections \ref{sec:Semi-analytic solution} and \ref{sec:Analytic solution}, respectively).
In the semi-analytic solution, the expression of $\beta_h$ in equation (\ref{eq:beta_h 2}) is used as it is, and is solved through numerical integration.
In the analytic solution, the above expression of $\beta_h$ is further approximated so that it is solved analytically (see Section \ref{sec:The approximated analytic jet head velocity}).

\subsection{The cocoon and jet collimation}
\label{sec:cocoon collimation}

\subsubsection{The system of equations}
We follow the same treatment of \citet{2011ApJ...740..100B}.
The unshocked jet's height $\hat{z}$ can be written as a function of the jet luminosity $L_j$ and the cocoon's pressure $P_c$:
\begin{eqnarray}
  \hat{z} = \sqrt{\frac{L_j}{\pi c P_c}} + z_*.
  \label{eq:z}
\end{eqnarray}
With $r_{in}$ being the radius at which the jet is injected into the medium, $z_* = \max[r_{in},z(P_c = P_{j0})]$ is the radius at which the pressure of the injected jet and the pressure of the cocoon are balanced; beyond $z_*$ the pressure of the cocoon is higher than the pressure of the injected jet (\citealt{2011ApJ...740..100B}).
In our simulations, $z_*$ is typically of the same order of $r_{in}$, hence, for simplicity, we take $z_* \approx r_{in}$.

At a certain time $t$, the jet is uncollimated if the jet head's radius, $r_h(t)$, is below $\hat{z}/2$, and collimated if it is beyond $\hat{z}/2$ (see Figure 2 in \citealt{2011ApJ...740..100B}). 
Hence, the jet head's cross-section can be found for the two modes as follows:
\begin{equation}
  \Sigma_j(t) =
    \begin{cases}
      \pi r_h^2(t)\theta_0^2 & \text{if $r_h(t)<\hat{z}/2$ (uncollimated jet)} ,\\
      \pi r_h^2(t)\theta_j^2(t) & \text{if $r_h(t)>\hat{z}/2$ (collimated jet)} ,
    \end{cases}       
    \label{eq:Sigma}
\end{equation}
where $\theta_0$ is the initial opening angle of the jet\footnote{The initial opening angle is given by $\theta_0 \approx \theta_{inj} + 1/\Gamma_0$ where $\theta_{inj}$ is the opening angle of the injected jet at $t=t_0$ and $r=r_{in}$, and $\Gamma_0$ is its initial Lorentz factor.}, and $\theta_j(t)$ is the opening angle of the jet head at a given time $t$. 

Since the cocoon shape is approximated to an ellipsoidal [see (vi) in Section \ref{sec:Main approximations}], $r_c(t)$ is the cocoon's lateral width at the radius $\frac{1}{2}r_h(t)$. 
$r_c(t)$ is determined by integrating the lateral velocity, $\beta_{\perp}$, with which the cocoon expands into the ambient medium at the radius $\sqrt{[r_h(t)/2]^2 + r_c^2(t)}\approx \frac{1}{2}r_h(t)$ [since $r_h(t)\gg r_c(t)$; see Figures \ref{fig:BNS case}, \ref{fig:collapsar case}, and \ref{fig:maps}]. 
At this radius, since the ambient medium's velocity $v_a(r_h/2)$ is $\leq v_m/2$, and considering the value of $v_m$ (see Table \ref{Table:sim}), $\Gamma_a(r_h/2)$ is $\approx 1$ and a non-relativistic treatment is reasonable.
$\beta_{\perp}$ is therefore determined by the ram pressure balance between the cocoon and the ambient medium at the radius $\frac{1}{2}r_h(t)$, giving:
\begin{eqnarray}
P_c \approx \rho_a(r_h/2,t) c^2 [\beta_{\perp} - \beta_{a,\perp}]^2 ,
\label{eq:jump and beta_perp}
\end{eqnarray}
where $\beta_{a,\perp}$ is the ambient medium's expansion velocity [see equation (\ref{eq:v_a})] in the lateral direction:
\begin{eqnarray}
\beta_{a,\perp} =\left[\frac{r_c(t)}{r_m(t)}\right]\frac{v_{m}}{c}.
\label{eq:beta_perp}
\end{eqnarray}
In summary, the equations describing the jet-cocoon system can be found as follows:
\begin{align}
\label{eq:S1}
  \frac{dr_c(t)}{dt} =& c\beta_\perp , \\
\label{eq:S2}
    \beta_{\perp} =&
    \sqrt{\frac{P_{c}}{{\rho}_{a}(r_h/2,t) c^{2}}} +\left[\frac{r_c(t)}{r_m(t)}\right]\frac{v_{m}}{c}, \\
    \label{eq:S3}
    P_{{c}} =& \:\:\:\:\:\: \frac{E_{i}}{3\:V_{{c}}} \:\:\:\:\:\:\:=  \eta\frac{L_j\left(1-\langle{\beta_h}\rangle \right) \:(t-t_0)}{2 \pi r_c^{2}(t) r_{{h}}(t)} , \\
    \label{eq:S4}
    \Sigma_j(t) =& \pi r_h^2(t) \theta_j^2(t) = \frac{L_j \theta_0^2}{4 c P_c} ,
\end{align}
where $P_c$ and $V_c$ are defined as in equations (\ref{eq:P_c}) and (\ref{eq:V_c}), respectively;
and $\left<\beta_h\right>=\frac{1}{c}\frac{r_h(t)-r_0}{t-t_0}$ is the time-averaged jet head velocity, which is a term that takes into account the fact that a part of the injected energy [$=L_j\langle{\beta_h}\rangle(t-t_0)$] is contained in the jet and does not make its way into the cocoon instantly. 
The last equation (\ref{eq:S4}) is determined by the pressure balance between the post-collimated jet and the cocoon (\citealt{2011ApJ...740..100B}).

The expression of $\beta_\perp$ [and eventually $r_c(t)=\int_{t_0}^t c \beta_\perp dt + r_0\theta_{0}$] here is different from the original 
collapsar case where the medium is static (\citealt{2011ApJ...740..100B}; \citealt{2018MNRAS.477.2128H}); it is instead applicable to both the case of static medium and the case of expanding medium. 
The term $\left[\frac{r_c(t)}{r_m(t)}\right]\frac{v_{m}}{c}$ in equation (\ref{eq:S2}) is new and is the result of the homologous expansion of the medium.
It is worth mentioning that the term $\left[\frac{r_c(t)}{r_m(t)}\right]\frac{v_{m}}{c}$ is far more dominant (in $\beta_\perp$) over the term $\sqrt{\frac{P_{c}}{\rho_{a}(r_h/2,t) c^{2}}}$ in the case of an expanding medium as in BNS mergers, and hence it is important.

\subsubsection{The parameters $\eta$ and $\eta'$}
\label{sec:eta definition}
$\eta$ in equation (\ref{eq:S3}) is a parameter that expresses the fraction of internal energy in the total energy delivered into the cocoon (by the engine and through the jet) at a given time $t$ (\citealt{2011ApJ...740..100B}; \citealt{2013ApJ...777..162M}). It takes values between 0 and 1, and it can be expressed as:
\begin{equation}
     \eta= \frac{3P_c V_c}{L_j\left(1-\langle{\beta_h}\rangle \right) \:(t-t_0)} ,
\label{eq:eta}
\end{equation}
with $E_i= 3P_cV_c$ [see equation (\ref{eq:P_c})]. 
For convenience, we define the parameter $\eta'=\eta[1-\langle{\beta_h}\rangle]$; 
it relates to the fraction of internal energy in the cocoon out of the total energy delivered by the central engine, at a given time $t$. 
Hence:
\begin{equation}
     \eta'=  \frac{3P_c V_c}{L_j \:(t-t_0)}.
\label{eq:eta'}
\end{equation}
$\eta$ and $\eta'$ can be easily deduced from numerical simulations by measuring both $P_c$ and $V_c$, or by measuring the internal energy in the cocoon $E_i$.
In Section \ref{sec:Mesurements of internal energy}, using numerical simulations' results, we will show that, on average, $\langle{\eta'}\rangle \sim 1/2$ for the collapsar case, and $\langle{\eta'}\rangle \sim 1/4$ for the BNS merger case [see Figure \ref{fig:eta} and equation (\ref{eq: eta cases})], where:
\begin{equation}
    \langle{\eta'}\rangle = \frac{1}{t_b-t_0}\int_{t_0}^{t_b}\eta'dt.
    \label{eq:eta average}
\end{equation}
These fiducial values will be adopted to solve the jet head motion (see Table \ref{Table:sim}).

\subsection{The semi-analytic solution}
\label{sec:Semi-analytic solution}
Here, the system of equations [equations (\ref{eq:L expression approx}), (\ref{eq:N_s}), (\ref{eq:beta_h 2}), (\ref{eq:z}), (\ref{eq:Sigma}), (\ref{eq:S1}), (\ref{eq:S2}), (\ref{eq:S3}), and (\ref{eq:S4})] is solved though numerical integration.
At every time step, the time is updated (from $t$ to $t+dt$, where $dt$ is sufficiently small).
The density ${\rho}_a(r_h/2,t)$ in equation (\ref{eq:S2}) is calculated using equation (\ref{eq:rho_a}).
Then, using equation (\ref{eq:S3}) the pressure is calculated; 
the parameter $\eta'$ [as defined in equation (\ref{eq:eta'})] is represented by its time-averaged value $\langle{\eta'}\rangle$ [see equation (\ref{eq:eta average})] as measured in numerical simulations (see Table \ref{Table:sim} for the values of $\langle{\eta'}\rangle$ used).
Next, $\beta_\perp$ is derived using equation (\ref{eq:S2}).
The jet head's cross-section and opening angle are found by calculating $\hat{z}$ first, using (\ref{eq:z}), and then determining the collimation mode and the opening angle of the jet, using equation (\ref{eq:Sigma}) together with equation (\ref{eq:S4}).
$\tilde{L}_c$ is then calculated using equations (\ref{eq:L expression approx}) and (\ref{eq:N_s}). 
Finally, at the end of each time step, the jet head radius $r_h(t)$, the cocoon's lateral width $r_c(t)$, and the cocoon's volume $V_c$, for the next time step are calculated using equations  (\ref{eq:beta_h 2}), (\ref{eq:S1}) and (\ref{eq:V_c}), respectively.
These processes are repeated until the jet breaks out of the ambient medium [i.e., the following condition is met: $r_h(t) \geqslant r_m(t)$].

\subsection{The analytic solution}
\label{sec:Analytic solution}
Here, the system of equations (\ref{eq:L expression approx}), (\ref{eq:N_s}), (\ref{eq:beta_h 2}), (\ref{eq:Sigma}), (\ref{eq:S1}), (\ref{eq:S2}), (\ref{eq:S3}), and (\ref{eq:S4}) [in Sections \ref{sec:Jump conditions}, \ref{sec:Main approximations}, and \ref{sec:cocoon collimation}] is simplified using several additional approximations, and then solved analytically.

In summary, the jet head's velocity [equation (\ref{eq:beta_h 2})] is simplified to equation (\ref{eq:beta_h 2 approx}), which can be written as a function of $t$, $r_h(t)$, and $\theta_j(t)$ using equations (\ref{eq:L expression approx}), (\ref{eq:N_s}), and (\ref{eq:Sigma}) [see Section \ref{sec:The approximated analytic jet head velocity}]. 
The expression of the cocoon's lateral width, $r_c(t)$, is simplified from equation (\ref{eq:S1}) to equation (\ref{eq:S1 approx}) [$\langle{\chi}\rangle$ can be found with equations (\ref{eq:chi}) and (\ref{eq:chi average}); see Section \ref{sec:The approximated cocoon's lateral width $r_c$}], and with equation (\ref{eq:S3 approx}) the expression of the cocoon pressure, $P_c$, is derived analytically in Section \ref{sec:The system of equations and the analytic solution} [in equation (\ref{eq:P_c approx}) as a function of $t$ and $r_h(t)$].
Next, equation (\ref{eq:S4}) is used to find the analytic expression of the jet opening angle $\theta_j(t)$ [equation (\ref{eq:theta_j/theta_0 app}) as a function of $r_h(t)$ and $t$], which allows us to derive an analytically solvable equation of motion of the jet head [equation (\ref{eq:dif dynamic coll v0})], and to determine the solution, $r_h(t)$, as a function of the initial parameters and $t$ (see Section \ref{sec:BNS merger case}).

The same logic can be used in the collapsar, and the equation of motion of the jet head can be found accordingly [equation (\ref{eq:dif static v0}); see Section \ref{sec:The collapsar case}].

For reference, Table \ref{Table:sim} presents a summary of the relevant parameters and the values they take.

\subsubsection{Approximated jet head velocity $\beta_h$}
\label{sec:The approximated analytic jet head velocity}
In the analytic solution, two additional approximations are used for the jet head velocity.
Firstly, in the case of BNS mergers where the medium is expanding (i.e., $\Gamma_a>1$), the term $\frac{1}{\Gamma_a^2}$ in the expression of $\tilde{L}_c$ in equation (\ref{eq:L expression approx}) is considered as constant and is absorbed into $N_s$. 
Secondly, in the analytic solution, the term $\left[ (1 - \beta_a)(1 + \tilde{L}_c^{1/2})^{-1} \right]$ in equation (\ref{eq:beta_h 2}) is also approximated as, roughly, constant over time
and is also effectively absorbed into the calibration coefficient $N_s$.
The result is the following expression:
\begin{eqnarray}
\beta_h  \approx \tilde{L}_c^\frac{1}{2} +  \beta_a .
\label{eq:beta_h 2 approx}
\end{eqnarray}
In the case of BNS mergers, and for typical parameters ($\beta_a \sim 0.2$ and ${\tilde L}_c \sim 0.1$--$0.4$), these approximations would result in a factor of $\sim 0.5$ being absorbed in $N_s$
[values of $N_s$ are given in the caption of Table \ref{Table:sim}; for details refer to Appendix \ref{sec:C2} and equation (\ref{eq:N_s analytic BNS})].
In the case of collapsars ($\beta_a = 0$) the above expression is even simpler [see equation (\ref{eq:beta_h 2 approx collapsar}) in Section \ref{sec:The collapsar case}], and this approximation results in a factor of $\sim 0.7$ being absorbed in $N_s$ [for details refer to Appendix \ref{sec:C2} and equation (\ref{eq:N_s analytic collaspar})].

\citet{2018MNRAS.477.2128H} showed that $N_s$ depends on the actual value of $\tilde{L}$ (i.e., $\tilde{L}_c$), but overall $N_s \sim 0.3-0.4$ for the case of a non-relativistic collapsar jet. 
As a remark, since $N_s$ here is used to absorb the above two approximations, its value differs depending on the type of the jet (BNS merger case or collapsar case) and on the type of the solution (semi-analytic or analytic; see Sections \ref{sec:Semi-analytic solution} and \ref{sec:Analytic solution}). 
Even for the case of a collapsar jet, the values of $N_s$ here do differ slightly from those in \citet{2018MNRAS.477.2128H}
[see the caption of Table \ref{Table:sim} for the values of $N_s$].
This is because additional difference in $N_s$ emerges as a result of the difference in the modeling [e.g., difference in the modeling of the cocoon's lateral width, volume, and in the value of $E_i/E_c$ (or $\eta$) compared to \citealt{2018MNRAS.477.2128H}].

\subsubsection{The approximated cocoon's lateral width $r_c$}
\label{sec:The approximated cocoon's lateral width $r_c$}
Here, the expression of $r_c(t)$ is simplified based on approximation that the term $\sqrt{\frac{P_{c}}{\rho_{a}(r_h/2,t) c^{2}}}$ in the expression of $\beta_\perp$ [in equation (\ref{eq:S2})] is considered as roughly constant over time.
This approximation is justified later by comparison with numerical simulations. 
This allows us to write equation (\ref{eq:S1}) as:
\begin{equation}
  \frac{dr_c(t)}{dt} + \left[-\frac{v_{m}}{r_m(t)}\right]r_c(t) = \sqrt{\frac{P_{c}}{\rho_{a}(r_h/2,t) }}.
\label{eq:S1 no jet}
\end{equation}
This is integrated, with $r_c(t=t_0)$ being very negligible, as
\begin{equation}
  r_c(t) \approx \chi(t) \sqrt{\frac{P_{c}}{\rho_{a}(r_h/2,t)}} (t-t_0),    
\end{equation}
where 
$\chi(t)$ is given by: 
\begin{equation}
  \chi(t) = \frac{r_m(t)}{r_m(t)-r_{m,0}} \ln{\left[\frac{r_m(t)}{r_{m,0}}\right]}\approx \frac{t-t_m}{t-t_0} \ln{\left[\frac{t-t_m}{t_0-t_m}\right]} ,
  \label{eq:chi}
\end{equation}
with $t_m$ being the time of the merger in the case of BNS mergers.
The value of $\chi(t)$ in BNS mergers depends on the time since the merger, the time since the jet launch, and the time delay between the merger and the jet launch. 
Typically $\chi(t)$ is found to take values as follows (also, see Table \ref{Table:sim} for the average values):
\begin{equation}
  \chi(t) 
    \begin{cases}
      =1 & \text{if $\beta_a=0$ (Collapsar case)} ,\\
      \sim 1-2 & \text{if $\beta_a \sim 0.2-0.3$ (BNS merger case)} .
    \end{cases}       
    \label{eq: chi cases}
\end{equation}

Since $\chi(t) \propto \ln{t}$, its evolution over time is very limited. Therefore, in order to further simplify the expression of $r_c(t)$, we consider the time-averaged value of $\chi(t)$:
\begin{equation}
    \langle{\chi}\rangle = \frac{1}{t_b-t_0}\int_{t_0}^{t_b}\chi(t)dt ,
    \label{eq:chi average}
\end{equation}
so that $r_c(t)$ is simplified to:
\begin{equation}
  r_c(t) \approx \langle{\chi}\rangle \sqrt{\frac{P_{c}}{\rho_{a}(r_h/2,t)}} (t-t_0).
\end{equation}
See Table \ref{Table:sim} for the typical values of $\langle{\chi}\rangle$.

When deriving the breakout time $t_b$, $\langle{\chi}\rangle$ and $t_b$ depend on each other [see equation (\ref{eq:t_b BNS}) for the expression of $t_b$, with $A_1 \propto 1/\sqrt{\langle{\chi}\rangle}$ as in equation (\ref{eq:A1})].
However, this dependency is very weak, and small variation in the value of $t_b$ hardly affects the value $\langle{\chi}\rangle$.
Therefore, both $t_b$ and $\langle{\chi}\rangle$ can be determined 
iteratively
\footnote{Initially, a typical value is assumed for $t_b$ and $\langle{\chi}\rangle$ based on a guess on $t_b$ which can be guided by numerical simulations. Then $t_b$ is found using equation (\ref{eq:t_b BNS}) and a new value of $\langle{\chi}\rangle$ is found by inserting $t_b$ in equation (\ref{eq:chi average}). This new value of $\langle{\chi}\rangle$ results in a slightly different $t_b$, which is used again (to find a more accurate $\langle{\chi}\rangle$). This process is repeated $\sim 2-3$ times until the values of $t_b$ and $\langle{\chi}\rangle$ converge.}.

\subsubsection{The system of equations and the analytic solution}
\label{sec:The system of equations and the analytic solution}
The system of equations (\ref{eq:S1}), (\ref{eq:S2}), (\ref{eq:S3}), and (\ref{eq:S4}) can be simplified to the following:
\begin{align}
\label{eq:S1 approx}
    r_c(t) \approx& \langle{\chi}\rangle \sqrt{\frac{P_{c}}{\rho_{a}(r_h/2,t)}} (t-t_0) ,\\
\label{eq:S2 approx}
    \beta_{\perp} =&
    \sqrt{\frac{P_{c}}{{\rho}_{a}(r_h/2,t) c^{2}}} +\left[\frac{r_c(t)}{r_m(t)}\right]\frac{v_{m}}{c}, \\
    \label{eq:S3 approx}
    P_{{c}} =& \:\:\:\:\:\: \frac{E_{i}}{3\:V_{{c}}} \:\:\:\:\:\:\:= \langle{\eta'}\rangle \frac{L_j \:(t-t_0)}{2 \pi r_c^{2}(t) r_{{h}}(t)} , \\
    \label{eq:S4 approx}
    \Sigma_j(t) =& \pi r_h^2(t) \theta_j^2(t) = \frac{L_j \theta_0^2}{4 c P_c} .
\end{align}
From equation (\ref{eq:rho_a}), ${\rho}_{a}(r_h/2,t)$ can be found.
Then, replacing $r_c(t)$ in equation (\ref{eq:S3 approx}), and using $r_h(t)\gg r_0$ and $r_{m,0}\gg r_0$, $P_c$ can be written as:
\begin{equation}
    P_c = \sqrt{ \frac{\langle{\eta'}\rangle}{\langle{\chi}\rangle^2} \frac{(3-n) L_j M_{a} }{2^{3-n}\: \pi^2(t-t_0)} \frac{r_m^{n-3}(t)}{r_h^{n+1}(t)}} .
    \label{eq:P_c approx}
\end{equation}
Finally, substituting equation (\ref{eq:P_c approx}) in equation (\ref{eq:S4 approx}) gives the expression of the opening angle of the collimated jet as:
\begin{equation}
    \frac{\theta_j(t)}{\theta_0} =
    \left[\frac{\langle{\chi}\rangle^2}{\langle{\eta'}\rangle} \frac{L_j}{M_{a}c^2} \frac{1}{(3-n)2^{n+1}}        \right]^{\frac{1}{4}}\: \left[\frac{r_h(t)}{r_m(t)} \right]^{\frac{n-3}{4}} \: [t-t_0]^{\frac{1}{4}} .
    \label{eq:theta_j/theta_0 app}
\end{equation}
Notice the weak dependence of the jet opening angle on time, which has already been pointed out in \citet{2020MNRAS.491.3192H}. 

The opening angle of the jet depends on the two parameters $\langle{\chi}\rangle$ and $\langle{\eta'}\rangle$. 
In the BNS merger case, the ratio $\langle{\chi}\rangle^2/\langle{\eta'}\rangle$ can take values up to $\sim 10$; hence, these two parameters are important and should not be overlooked.

The expression of $P_c$ and $\theta_j(t)$ [equations (\ref{eq:P_c approx}) and (\ref{eq:theta_j/theta_0 app})] is valid for both a BNS merger jet case and a collapsar jet case [where, $r_m(t)\equiv r_m$ and $\langle{\chi}\rangle=1$].

\subsubsection{Analytic solution for the BNS merger case}
\label{sec:BNS merger case}

The jet head velocity [equation (\ref{eq:beta_h 2 approx})] with equations (\ref{eq:L expression approx}), (\ref{eq:v_a}), (\ref{eq:rho_a}), (\ref{eq:N_s}), and (\ref{eq:theta_j/theta_0 app}), with further simplifications, gives the following differential equation:
\begin{eqnarray}
    \frac{dr_h(t)}{dt}  + \left[ -\frac{v_{m}}{r_{m}(t)}\right]r_h(t) = {A(t)}\:{r_m(t)}^\frac{3-n}{2}{r_h(t)}^\frac{n-2}{2} ,
    \label{eq:dif dynamic coll v0}
\end{eqnarray}
where:
\begin{equation}
A(t)= N_s\sqrt{ \left(\frac{r_{m,0}^{3-n}-r_0^{3-n}}{(3-n)\:r_{m,0}^{3-n}}\right)\left(\frac{4\:L_j}{\theta_0^2 M_{a}\:c}\right)    } \times\left[\frac{\theta_0}{\theta_j(t)}\right] .
\label{eq:A(t)}
\end{equation}
The jet is initially uncollimated until the jet head's radius, $r_h(t)$, crosses the radius $\hat{z}/2$ [see equation (\ref{eq:Sigma})].
Since this initial phase is very short (relative to the jet propagation timescale until the breakout; see Figure \ref{fig:BNS case}), we consider as if the jet is in the collimated mode from the start ($t=t_0$).
Therefore, the jet opening angle can be found using equation (\ref{eq:theta_j/theta_0 app}), and after inserting it in the expression of $A(t)$ [equation (\ref{eq:A(t)})], the equation of motion [equation (\ref{eq:dif dynamic coll v0})] can be found as:
\begin{equation}
    \left[\frac{r_h(t)}{r_m(t)}\right]^{\frac{5-n}{4}} =  A_1 \frac{5-n}{4} \int {r_m^{-3/4}(t)[1-r_{m,0}/r_m(t)]^{-1/4} dt},
\end{equation}
where $A_1$, a constant, can be found as follows:
\begin{align}
\label{eq:A1}
A_1=& N_s \left[\frac{\langle{\eta'}\rangle}{\langle{\chi}\rangle^2}\right]^\frac{1}{4}\left[ \left(\frac{r_{m,0}^{3-n}-r_0^{3-n}}{r_{m,0}^{3-n}}\right)^2\frac{2^{n+5}}{3-n}\frac{\:L_j\:v_{m}}{\theta_0^4 M_{a}}        \right]^{\frac{1}{4}} , \\
\label{eq:A1 approx}
 \approx& N_s \left[\frac{\langle{\eta'}\rangle}{\langle{\chi}\rangle^2}\right]^\frac{1}{4}\left[\frac{2^{n+5}}{3-n}\frac{\:L_j\:v_{m}}{\theta_0^4 M_{a}}        \right]^{\frac{1}{4}} .
\end{align}
In the case where the delay between the merger time and the jet launch time, $t_0-t_m$, is significantly smaller in comparison to the breakout time $t_b-t_m$:
$t_0-t_m\ll t_b - t_m$, we have $r_m(t) \gg r_{m,0}$, hence the following approximation can be made:\footnote{The approximation $[1-r_{m,0}/r_m(t)]\approx 1$ is not good in the early phase of jet propagation where $r_{m,0} \simeq r_m(t)$. 
Still, since this early phase's timescale is very short (relative to the whole jet propagation timescale; see Figure \ref{fig:BNS case}) this approximation is reasonable as long as $t_b-t_m\gg t_0-t_m$.}
\begin{equation}
    \int {r_m^{-3/4}(t)[1-r_{m,0}/r_m(t)]^{-1/4} dt}  \simeq \int{ r_m^{-3/4}(t)dt}.
    \label{eq:r_m>>r_m,0 app}
\end{equation}
With the boundary conditions $r_m(t_0)=r_{m,0}$ and $r_h(t_0)=r_0$ at $t=t_0$, the integration gives:
\begin{equation}
    r_h(t) = \left\{ \frac{(5-n)A_1 }{v_{m}}(r_m^{\frac{1}{4}}(t) - r_{m,0}^\frac{1}{4}) +\left[\frac{r_0}{r_{m,0}}\right]^\frac{5-n}{4} \right\}^\frac{4}{5-n}r_m(t) .
    \label{eq:r_h analytic}
\end{equation}
The jet head velocity can be deduced from equation (\ref{eq:r_h analytic}) as:
\begin{equation}
v_h(t)=v_{m}\left[ \frac{r_h(t)}{r_m(t)}\right] +A(t)\left[ \frac{r_h(t)}{r_m(t)}\right]^\frac{n-2}{2}[r_m(t)^\frac{1}{4}(r_m(t)-r_{m,0})^\frac{1}{4}] .
\label{eq:v_h BNS}
\end{equation}
Finally, the breakout time can be derived by taking $r_h(t_b)/r_m(t_b)=1$ in equation (\ref{eq:r_h analytic}):
\begin{equation}
t_b - t_0 =\left\{ \frac{v_{m}^\frac{3}{4}}{(5-n)A_1}\left[1-\left[\frac{r_0}{r_{m,0}}\right]^\frac{5-n}{4}\right]+ \left(\frac{r_{m,0}}{v_m}\right)^\frac{1}{4}\right\}^{4} - \frac{r_{m,0}}{v_{m}} .
\label{eq:t_b BNS}
\end{equation}

\subsubsection{Analytic solution for the collapsar case}
\label{sec:The collapsar case}
This case is a special case from the previous one (in Section \ref{sec:BNS merger case}) where $v_m=0$ [i.e., the ambient medium is static: $\beta_a = 0$, $\chi(t)=1$ and $r_m(t) \equiv r_{m}$]. 
Therefore, the equation of motion [equation (\ref{eq:beta_h 2})], after being approximated to equation (\ref{eq:beta_h 2 approx}) (see Section \ref{sec:The approximated analytic jet head velocity}),
can be further simplified to the following:
\begin{eqnarray}
\beta_h  \approx \tilde{L}_c^\frac{1}{2}.
\label{eq:beta_h 2 approx collapsar}
\end{eqnarray}
Hence, the equation of motion for the jet head can be found as:
\begin{eqnarray}
    \frac{dr_h(t)}{dt}  = {A(t)}\:{r_m}^\frac{3-n}{2}{r_h(t)}^\frac{n-2}{2} ,
    \label{eq:dif static v0}
\end{eqnarray}
where $A(t)$ here is:
\begin{equation}
A(t)= N_s\sqrt{ \left(\frac{r_{m}^{3-n}-r_0^{3-n}}{(3-n)\:r_{m}^{3-n}}\right)\left(\frac{4\:L_j}{\theta_0^2 M_{a}\:c}\right)    } \times\left[\frac{\theta_0}{\theta_j(t)}\right] ,
\label{eq:A(t) cc}
\end{equation}
which is the same expression as in equation (\ref{eq:A(t)}) [where here $r_{m,0} \equiv r_{m}$].
As in Section \ref{sec:BNS merger case}, the initial uncollimated phase is overlooked for simplicity.
Then the expression of $\theta_j(t)/\theta_0 $ can be found using equation (\ref{eq:theta_j/theta_0 app}) [with $r_
{m}(t)=r_m$ in the collapsar case].
Inserting $\theta_j(t)/\theta_0 $ in the above expression of $A(t)$ [equation (\ref{eq:A(t) cc})], 
integrating equation (\ref{eq:dif static v0}), and using the boundary condition $r_h(t_0)=r_0$, gives the following expression for the jet head radius:
\begin{equation}
r_h(t) = \left\{\frac{(5-n)A_1'}{3} (t - t_0)^\frac{3}{4} +\left[\frac{r_0}{r_{m}}\right]^\frac{5-n}{4} \right\}^\frac{4}{5-n} r_m,
    \label{eq:r_h collpsar}
\end{equation}
and the jet head velocity:
\begin{equation}
v_h(t) = A_1'r_m\left\{\frac{(5-n)A_1'}{3} (t - t_0)^\frac{3}{4} +\left[\frac{r_0}{r_{m}}\right]^\frac{5-n}{4} \right\}^\frac{n-1}{5-n}(t-t_0)^{-\frac{1}{4}},
    \label{eq:v_h collapsar}
\end{equation}
where $A_1'$ is a constant that can be written as:
\begin{align}
\label{eq:A1'}
A_1'=& N_s\left[\frac{\langle{\eta'}\rangle}{\langle{\chi}\rangle^2}\right]^\frac{1}{4}\left[ \left(\frac{r_{m}^{3-n}-r_0^{3-n}}{r_{m}^{4-n}}\right)^2\frac{2^{n+5}}{3-n}\frac{L_j}{\theta_0^4 M_{a}}        \right]^{\frac{1}{4}} , \\
\label{eq:A1' approx}
\approx& N_s\left[\frac{\langle{\eta'}\rangle}{\langle{\chi}\rangle^2}\right]^\frac{1}{4}\left[ \frac{1}{r_{m}^{2}}\frac{2^{n+5}}{3-n}\frac{L_j}{\theta_0^4 M_{a}}        \right]^{\frac{1}{4}} .
\end{align}
The breakout time can be found for $r_h(t_b)=r_m$ as:
\begin{equation}
t_b - t_0 = \left\{ \frac{3}{(5-n)A_1'}\left[ 1-\left(\frac{r_0}{r_m}\right)^\frac{5-n}{4}\right]\right\}^\frac{4}{3}.
\label{eq:t_b collapsar}
\end{equation}

\section{Comparison with numerical simulations}
\label{sec:3}
\subsection{Numerical simulations}
In addition to the analytic (and semi-analytic) modeling presented above, we carried out a series of 2D relativistic hydrodynamical simulations. 
In total, the series includes a total of over a hundred models covering a wide parameter space (see Table 1 in \citealt{2020MNRAS.491.3192H}).
The essential aim of carrying out numerical simulations here is to test the semi-analytic (Section \ref{sec:Semi-analytic solution}) and the analytic (Section \ref{sec:Analytic solution}) solutions, and calibrate them if necessary.
These tests and calibrations are presented for both, the case of BNS merger jet, and the case of collapsar jet.

We pick up four models, as a subsample, out of our sample of numerical simulations.
As presented in Table \ref{Table:sim}, two out of four are BNS merger models, with different initial opening angles (T03-H and T13-H); and the other two are collapsar models, with different initial opening angles as well (A and B). 
The parameters of the stellar envelope in collapsar simulations (models A and B) follows the widely used model 16TI in \citet{2006ApJ...637..914W}. 
However, for simplicity, the density profile is approximated to a power-law function with an index $n=2$ [see (iv) in Section \ref{sec:Main approximations}].
This allows the analytic results to be tested fairly with simulations.

The injection radius, $r_{in}$, is set at $1.2\times 10^8$ cm for the BNS merger case, and $10^9$ cm for the collapsar case (see Table \ref{Table:sim}). 
This might seem quite large; 
ideally the injection radius should be of the order of $10^7$ cm.
However, since the density profile in the inner region can be approximated to a power-law function with an index $n<3$ ($n=2$ for the dynamical ejecta of BNS mergers, see Figure 8 in \citealt{2020MNRAS.491.3192H}; $n\approx 1.5$ for the collapsar case, see Figure 2 in \citealt{2013MNRAS.428..729M}), 
the mass contained in the inner region ($<10^8$ cm in BNS mergers; $<10^9$ cm in collapsars) is negligible, in comparison to the total ambient medium mass, as long as $r_{in} \ll r_{m,0}$ (or $r_{in} \ll r_{m}$), which is the case in our simulations [see equation (\ref{eq:M_a})]. 
Hence, this inner region is expected to have a very limited effect on the overall jet dynamics.
For an estimation of the effect of these values on jet dynamics, please refer to the analytic model [in particular refer to equations (\ref{eq:t_b BNS}), and (\ref{eq:t_b collapsar}), for the effect of the value of the inner boundary, $r_0$, on the jet breakout time]. 
Note that the motivation for taking such large values for $r_{in}$ is because smaller injection radii make numerical simulation extremely expensive in terms of computation time.  

Further details about the numerical code are presented in \citet{2017MNRAS.469.2361H}. 
For more information about the setup of the numerical simulations, refer to Section 3.1 in \citet{2020MNRAS.491.3192H}. 

\subsection{Measurement of the internal energy in the cocoon and $\eta'$}
\label{sec:Mesurements of internal energy}
\begin{figure}
    \vspace{4ex}
  \begin{subfigure}
    \centering
    \includegraphics[width=0.995\linewidth]{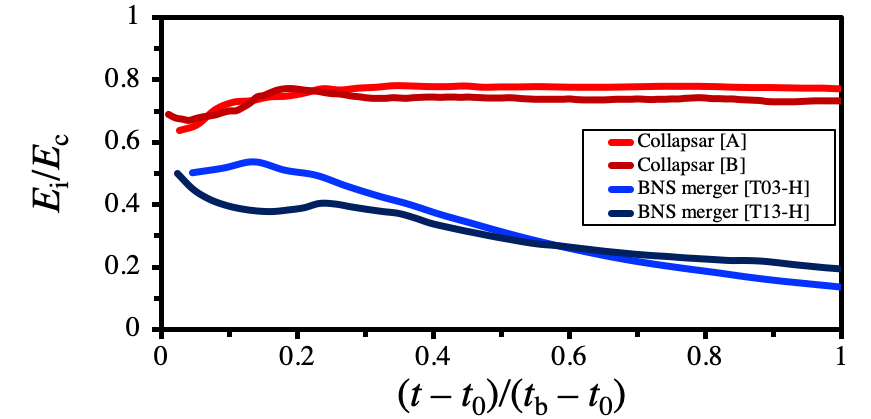} 
  \end{subfigure}
  \begin{subfigure}
    \centering
    \includegraphics[width=0.995\linewidth]{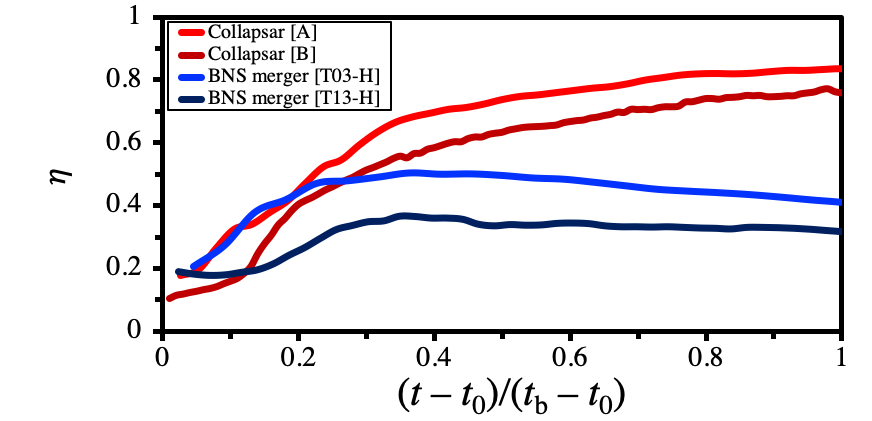} 
  \end{subfigure} 
  \begin{subfigure}
    \centering
    \includegraphics[width=0.995\linewidth]{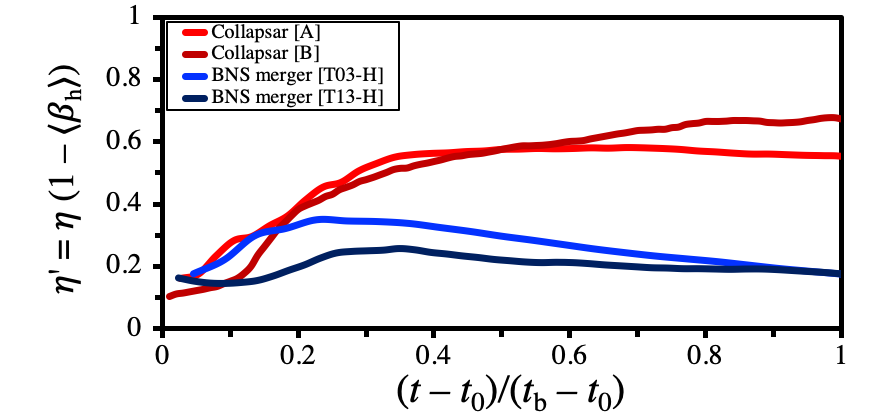} 
  \end{subfigure} 
  \caption{The fraction of the internal energy to the total energy of the cocoon (top); and the parameters $\eta$ (middle; the fraction of the cocoon's internal energy to the energy injected into the cocoon) and $\eta'$ (bottom; the fraction of the cocoon's internal energy to the injected jet energy), as measured in our 2D simulations [see equations (\ref{eq:eta}) and (\ref{eq:eta'})]. The red and dark red lines are for collapsar jet models (models A and B in Table \ref{Table:sim}). The blue and dark blue lines are for BNS merger models (models T03-H and T13-H in Table \ref{Table:sim}).}
  \label{fig:eta} 
\end{figure}

Figure \ref{fig:eta} shows the time evolution of the fraction of internal energy in the cocoon $E_i/E_c$, and the two parameters $\eta$ and $\eta'$ [previously defined in equations (\ref{eq:eta}) and (\ref{eq:eta'}); see Section \ref{sec:eta definition}], from the jet launch time ($t=t_0$) to the jet breakout time ($t=t_b$), as measured from numerical simulation
\footnote{The data was deduced by measuring $E_i$ (or the cocoon pressure $P_c$ as presented in Figure \ref{fig:P_c}, combined with the total volume of the cocoon $V_c$), the total energy contained in the cocoon $E_c$, and the jet head radius [$r_h(t)$ or $\langle{\beta_h}\rangle$],  all from numerical simulations.}.
For comparison, both the case of collapsar jet and the case of BNS merger jet are shown.
In the collapsar jet case (model A and B), the fraction of internal energy in the cocoon, $E_i/E_c$, is high ($\sim 0.7 - 0.8$).
The values of $\eta$ and $\eta'$ (in the range $\sim 0.5 - 1$) are also high. 
On the other hand, in the BNS merger jet case (model T03-H and T13-H), where the medium is expanding, the values of $E_i/E_c$, $\eta$, and $\eta'$ are significantly lower ($< 0.5$). 

This contrast is related to the adiabatic expansion of the cocoon, which is very effective in the case where the medium is expanding;
as the medium's expansion velocity (i.e., the dynamical ejecta's radial velocity) is comparable to the jet head velocity, up to the breakout time, this expansion enhances the volume of the (over-pressurized) cocoon significantly, while depleting its internal energy.
Through this process, in the BNS merger case, the inner region of the cocoon, 
where initially the density is high and the expansion velocity is small, 
is propelled further outward within the cocoon, up to velocities of the order of the homologous expansion of the medium, gaining kinetic energy at the expense of internal energy.
Note that in the collapsar case, the same process happens but to a lesser extent; 
the inner cocoon is propelled outward, but as a result of the initially static medium,
the gained velocity (and fraction of kinetic energy) is much less significant. 

Another reason is the high jet head velocity ($\langle{\beta_h}\rangle$) in the BNS merger jet case (roughly $ \sim 2 v_m/c$; see \citealt{2018PTEP.2018d3E02I}), implying that the fraction of the injected energy that reaches the cocoon [$\propto (1-\langle{\beta_h}\rangle)$] is less significant [relative to the case of a collapsar jet; see equation (\ref{eq:E_in})].
For more details refer to Appendix \ref{sec:A}.

To our best knowledge, this is the first time that the fraction of internal energy of the cocoon, and the parameter $\eta$ (and $\eta'$), has been measured (from simulations), and such a significant difference, between the case of collapsar jet and the case of BNS merger jet, has been found.
The parameter $\eta$ has been discussed in the literature in the case of collapsar jets, and it has been suggested to take a value of $\sim 1$ (e.g., in \citealt{2011ApJ...740..100B}). 
Our results show that this assumption is quite reasonable.
On the other hand,
several recent works naively assumed the same value, $\eta \sim 1$, for the case of BNS merger jet (e.g., \citealt{2018ApJ...866L..16M}; \citealt{2019ApJ...876..139G}; \citealt{2020A&A...636A.105S}). 
Here, we show that such assumption is not reasonable
by a factor of $\sim 2$; $\eta$ should rather be smaller in the case of BNS merger jet, as it can be seen in Figure \ref{fig:eta} (unless $t_b -t_m \sim t_0 - t_m$). 

In summary, $\eta'$ is found to take values as follows:
\begin{equation}
  \eta'  
    \begin{cases}
      \sim 0.5-1 & \text{if $\beta_{a} = 0$ (Collapsar jet case)} ,\\
      \sim 0.1-0.5 & \text{if $\beta_{a} \gg 0$ (BNS merger jet case)} .
    \end{cases}       
    \label{eq: eta cases}
\end{equation}
As a remark, for typical cases, naively assuming $\eta \sim 1$ for the case of BNS merger tends to incorrectly give a factor of $\sim \sqrt{2}$ times more collimation [i.e., $\sim \sqrt{2}$ times more jet head velocity, and hence much shorter breakout times; see equations (\ref{eq:theta_j/theta_0 app}); (\ref{eq:A1}); and (\ref{eq:t_b BNS})].

\begin{figure*}
    \vspace{4ex}
  \begin{subfigure}
    \centering
    \includegraphics[width=0.495\linewidth]{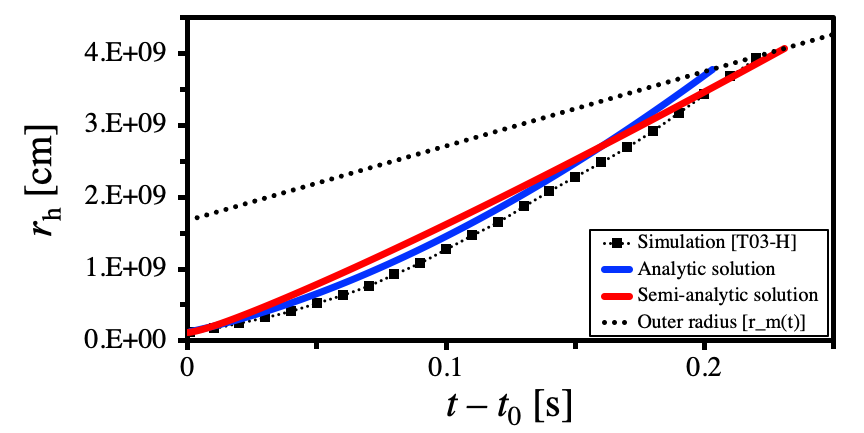}
  \end{subfigure}
  \begin{subfigure}
    \centering
    \includegraphics[width=0.495\linewidth]{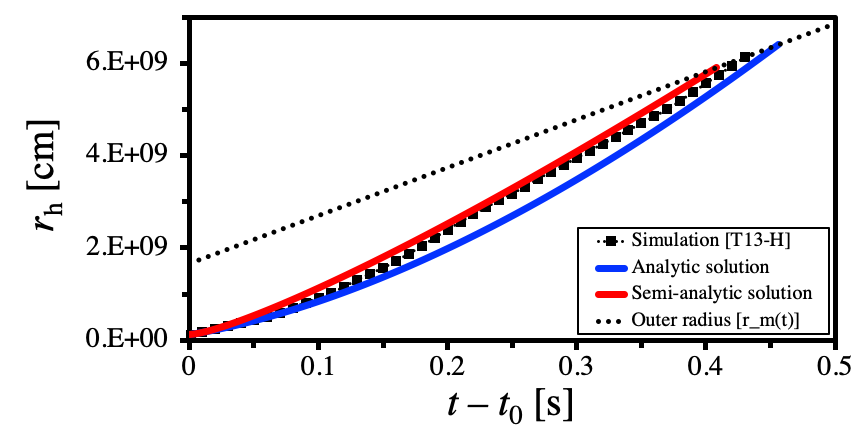} 
  \end{subfigure} 
    \begin{subfigure}
    \centering
    \includegraphics[width=0.495\linewidth]{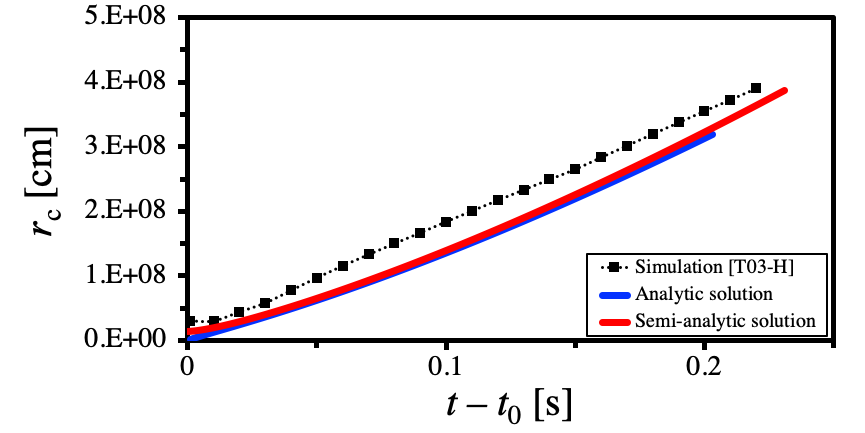} 
  \end{subfigure}
  \begin{subfigure}
    \centering
    \includegraphics[width=0.495\linewidth]{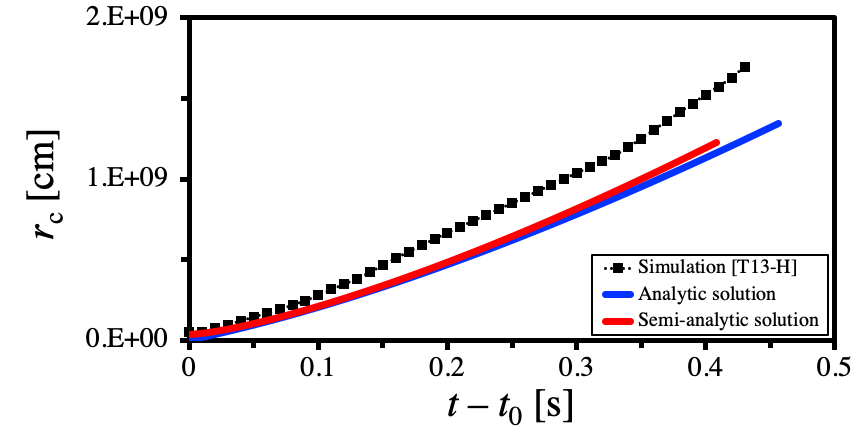} 
  \end{subfigure}
  \begin{subfigure}
    \centering
    \includegraphics[width=0.495\linewidth]{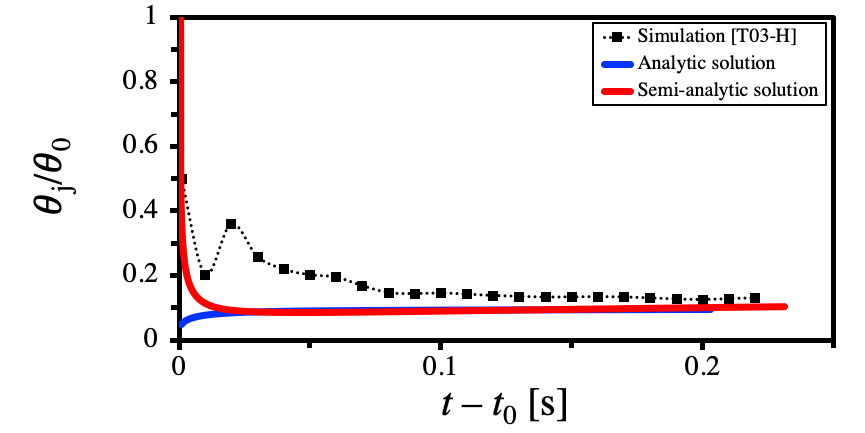} 
  \end{subfigure}
  \begin{subfigure}
    \centering
    \includegraphics[width=0.495\linewidth]{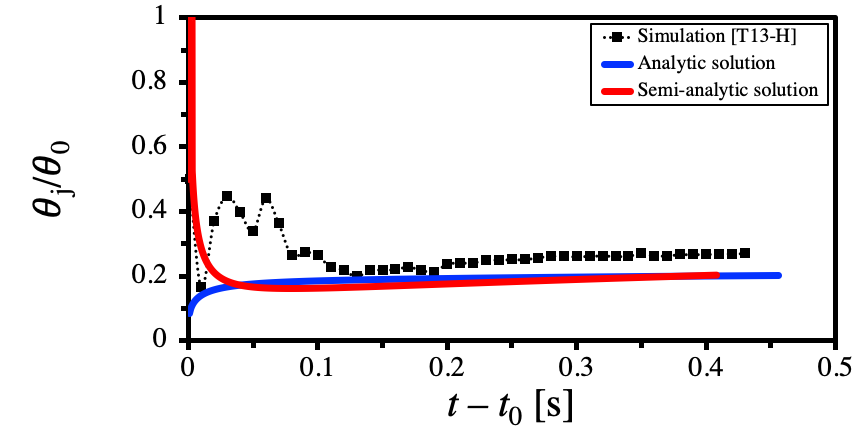} 
  \end{subfigure}
  \caption{Results for the case of BNS mergers showing the jet's and the cocoon's evolution over time, as measured in numerical simulations (black dotted line with filled squares), and as inferred with the analytic (solid blue line) and the semi-analytic (solid red line) solutions [see Sections \ref{sec:Semi-analytic solution} and \ref{sec:BNS merger case}]. 
  The black dotted line in the top two panels shows the outer radius of the expanding ejecta.
  From top to bottom, the jet head radius, the cocoon's lateral width, and the jet opening angle relative to the initial opening angle, are shown respectively. From left to right, results for the narrow jet case (model T03-H) and the wide jet case (model T13-H) are shown respectively. We take $N_s=0.46$ in the analytic solution, and $N_s=0.75$ in the semi-analytic solution (see Section \ref{sec:The approximated analytic jet head velocity} and Table \ref{Table:sim}).}
  \label{fig:BNS case} 
\end{figure*}

\begin{figure*}
    \vspace{4ex}
  \begin{subfigure}
    \centering
    \includegraphics[width=0.494\linewidth]{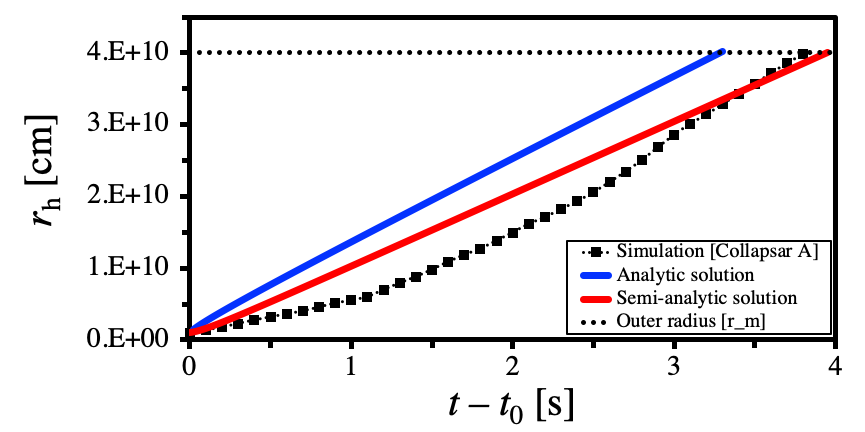} 
  \end{subfigure}
  \begin{subfigure}
    \centering
    \includegraphics[width=0.494\linewidth]{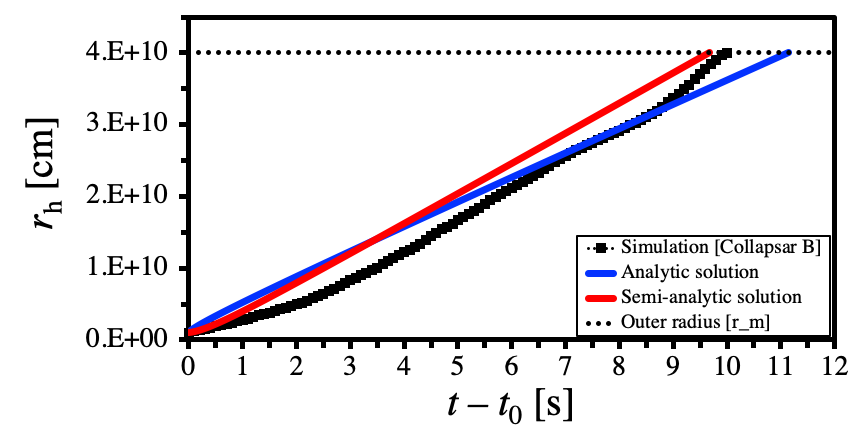} 
  \end{subfigure} 
    \begin{subfigure}
    \centering
    \includegraphics[width=0.494\linewidth]{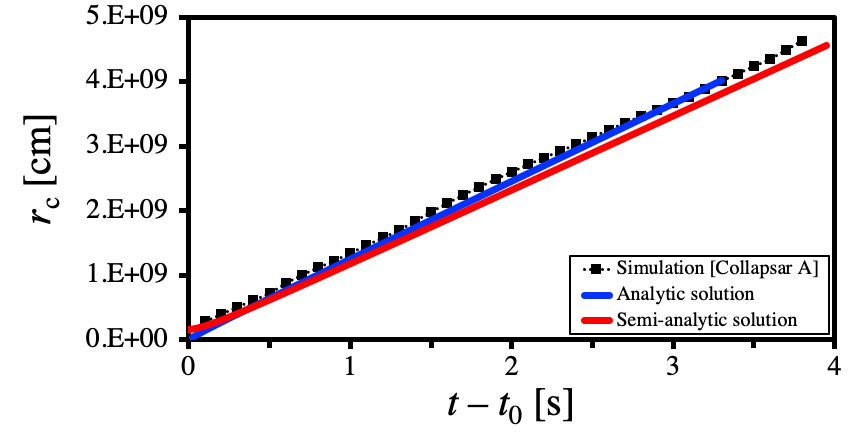} 
  \end{subfigure}
  \begin{subfigure}
    \centering
    \includegraphics[width=0.494\linewidth]{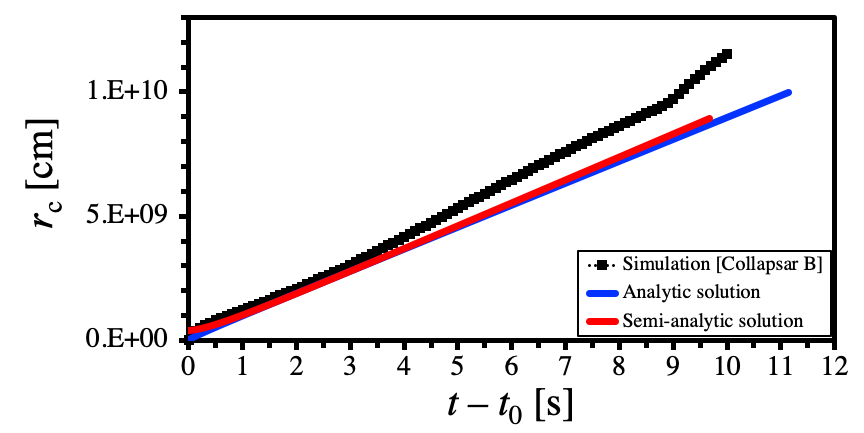} 
  \end{subfigure} 
  \begin{subfigure}
    \centering
    \includegraphics[width=0.494\linewidth]{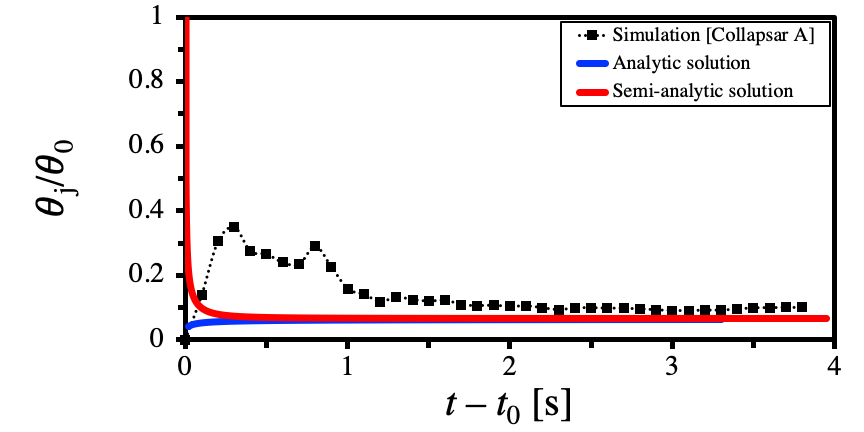} 
  \end{subfigure} 
  \begin{subfigure}
    \centering
    \includegraphics[width=0.494\linewidth]{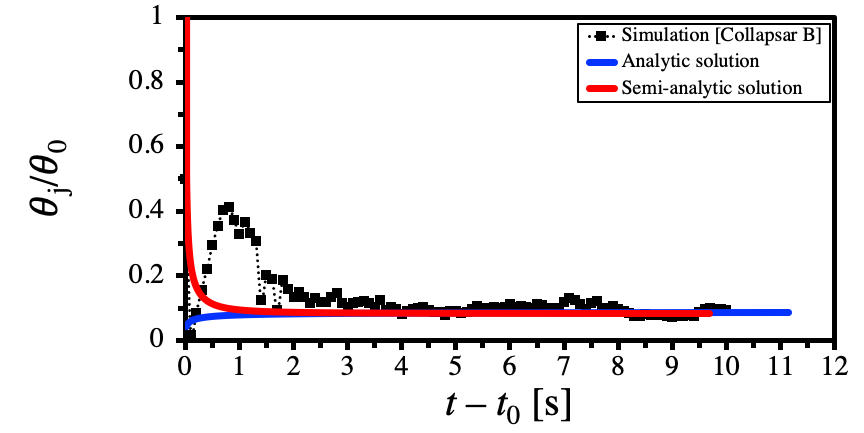} 
  \end{subfigure} 
  \caption{Same as Figure \ref{fig:BNS case} for the collapsar case, showing the results from numerical simulations (black dotted line with filled squares), and results of the analytic (solid blue line) and the semi-analytic (solid red line) solutions [see Sections \ref{sec:Semi-analytic solution} and \ref{sec:The collapsar case}]. 
  The horizontal black dotted line in the top two panels shows the radius of the stellar envelope.
  From left to right, the results for the narrow jet case (model A) and the results for the wide jet case (model B) are shown, respectively. We take $N_s=0.38$ in the analytic solution, and $N_s=0.53$ in the semi-analytic solution (see Section \ref{sec:The approximated analytic jet head velocity} and Table \ref{Table:sim}).
  }
  \label{fig:collapsar case} 
\end{figure*}

\begin{table*}
\caption {A subsample showing the simulated models and the corresponding parameters. 
From the left: 
The model name; 
the type of jet (BNS merger or collapsar); 
the ambient medium's mass; 
the jet initial opening angle;  
the engine's isotropic equivalent luminosity [$L_{iso,0} = \frac{2 L_j}{1-\cos\theta_0}\simeq \frac{4 L_j}{\theta_0^2}$] where $L_j$ is the jet true luminosity (one sided); 
the inner radius at which the jet is injected in simulations; 
the ambient medium's outer radius at the start of the simulation; 
the maximum expansion velocity of the ambient medium; 
the time-averaged value of $\eta'$ [see equation (\ref{eq:eta average})] estimated from simulations (see Figure \ref{fig:eta}); 
the time-averaged value of $\chi(t)$ used in the analytic solution [using equation (\ref{eq:chi average}); see Section \ref{sec:The approximated cocoon's lateral width $r_c$}];
the breakout time measured in numerical simulations;
the inferred breakout time by the analytic solution [using equation (\ref{eq:t_b BNS}) for the BNS merger jet case, and equation (\ref{eq:t_b collapsar}) for the collapsar jet case], and by the semi-analytic solution (see Section \ref{sec:Semi-analytic solution}); 
The values of the calibration coefficient $N_s$ are: 
$0.46$ and $0.75$ in the BNS merger case (for the analytic and the semi-analytic solution, respectively), $0.38$ and $0.53$ in the collapsar case (for the analytic and the semi-analytic solution, respectively).
The density profile of the ambient medium in all models is approximated to power-law with the index $n=2$ [see (iv) in Section \ref{sec:Main approximations}].}
\label{Table:sim}
\begin{tabular}{llllllllll|l|l|l}
  \hline
        &  & $M_{a}$  & $\theta_0$ &  $L_{iso,0}$ & $r_{in}$ & $r_m(t_0)$ & $v_{m}$   & $\langle{\eta'}\rangle$  & $\langle{\chi}\rangle$ & $t_b-t_0$ [s] & $t_b-t_0$ [s] & $t_b-t_0$ [s] \\
        &  & & & & & & & & &  &  & (Semi-\\
    Model & Type &  [$M_\odot$] &  [deg] & [erg s$^{-1}$] & [cm] & [cm] &  [c]  & & & (Simulation) & (Analytic) & analytic)\\
        \hline
    T03-H & BNS  & $0.002$ & 6.8 & $5\times10^{50}$ & $1.2\times 10^8$ & $1.67\times10^9$ & $\frac{\sqrt{3}}{5}$ & $1/4$ & $1.25$ & 0.221 & 0.203 & 0.231\\
    T13-H & BNS  & $0.002$ & 18.0 & $5\times10^{50}$ & $1.2\times 10^8$ & $1.67\times10^9$ & $\frac{\sqrt{3}}{5}$ & $1/4$ & $1.48$ & 0.429 & 0.456 & 0.408\\
         \hline
    A & Collapsar & $13.950$ & 9.2 & $7.83\times10^{52}$ & $10^9$ & $4\times10^{10}$ & 0 & $1/2$ & $1.00$ & 3.804 & 3.282 & 3.947\\
    B & Collapsar & $13.950$ & 22.9 & $1.27\times10^{52}$ & $10^9$ & $4\times10^{10}$ & 0 & $1/2$ & $1.00$ & 9.930 & 11.137 & 9.681 \\
        \hline
  \hline
 \end{tabular}
\end{table*}
 
\subsection{Time evolution of the jet propagation}
\label{sec:Time evolution of the jet propagation}
\subsubsection{BNS merger's case}
\label{BNS merger's case}
In Figure \ref{fig:BNS case} we show results for the two models of jet propagation in the BNS merger ejecta [T03-H and T13-H with $\theta_0=6.8^\circ$ (left) and $18.0^\circ$ (right), respectively; see Table \ref{Table:sim}].
Three different quantities are shown (from top to bottom): the jet head radius $r_h(t)$, the cocoon's lateral width $r_c(t)$, and the opening angle of the jet head $\theta_j(t)$.

The calibration coefficient $N_s$ has been used to calibrate the analytic solution (with $N_s = 0.46$) and the semi-analytic solution (with $N_s = 0.75$); the value of $N_s$ is set so that the breakout time in the analytic (or semi-analytic) solution matches the breakout time in numerical simulations [refer to equation (\ref{eq:N_s}) and the explanation that follows]. 
Also, as noted in Section \ref{sec:The approximated analytic jet head velocity}, the different values of $N_s$ are due to the additional approximations in the analytic solution. 
It should be noted that the value of $N_s$ for the analytic model, here (0.46), is slightly different from the one in the analytic model presented in \citet{2020MNRAS.491.3192H} (0.40).
This difference is due to the main difference between the two models;
in \citet{2020MNRAS.491.3192H} the jet opening angle is fixed using the parameter $f_j$ (measured from numerical simulations; see Figure 3 in \citealt{2020MNRAS.491.3192H}), 
while here, the jet opening angle is determined self-consistently (by calculating the jet collimation by the cocoon), and should be more robust (see the bottom two panels in Figure \ref{fig:BNS case}).
For more details on $N_s$ refer to Appendix \ref{sec:C}.

The time evolution of the analytic and the semi-analytic jet head radius, $r_h(t)$, in Figure \ref{fig:BNS case} shows a very good agreement with simulations (within $\sim10\%$). 
Analytic and semi-analytic results hold fairly well for both models (T03-H and T13-H) showing that the jet-cocoon model here works well regardless of the initial jet opening angle $\theta_0$.

The time evolution of the analytic and semi-analytic cocoon's lateral width $r_c(t)$ is also consistent with simulations, especially for the case of small $\theta_0$ (within $\sim 10 \%$; see T03-H in Figure \ref{fig:BNS case}).
For the case of large $\theta_0$ (T13-H), the agreement with simulations is less significant, but roughly within $\sim 30 \%$, where $r_c(t)$ is slightly underestimated in the analytic and semi-analytical model (relative to numerical simulations).

In simulations, the jet head's opening angle has been estimated by taking the average opening angle from $r = \frac{1}{2}r_h(t)$ to $r = r_h(t)$ [see equation (34) in \citealt{2020MNRAS.491.3192H}].
This average opening angle is compared with the analytic and the semi-analytic jet opening angles in Figure \ref{fig:BNS case}.
Except the early time evolution of the jet opening angle, during which the jet-cocoon is highly inhomogeneous in simulations [in particular, in terms of entropy and Lorentz factor which are used to discriminate the jet from the cocoon: for more details see Section 3.1.2 in \citet{2020MNRAS.491.3192H}],
analytic and semi-analytic jet opening angles are consistent with the average opening angle in simulations within 
$\sim 30\%$.

\subsubsection{Collapsar's case}
\label{Collapsar's case}
Figure \ref{fig:collapsar case} shows a comparison of the analytic and the semi-analytic results with simulations for three quantities $r_h(t)$, $r_c(t)$, and $\theta_j(t)$, in the same manner as in Figure \ref{fig:BNS case} (and Section \ref{BNS merger's case}), but for the collapsar jet case.
We present two models with different initial opening angles [A (left) and B (right), with $\theta_0=9.2^\circ$ and $22.9^\circ$, respectively; see Table \ref{Table:sim}].
The calibration coefficient is found as $N_s = 0.38$ for the analytic solution, and $N_s = 0.53$ for the semi-analytic solution (see Section \ref{sec:The approximated analytic jet head velocity} for more details about the origin of this difference).
Although slightly larger, these values of $N_s$ are fairly consistent with those found by \citet{2018MNRAS.477.2128H}, despite several differences in the jet-cocoon modeling (such as for the cocoon's lateral width, volume, $\langle{\eta'}\rangle$, etc.). 

The analytic and the semi-analytic solutions for the time evolution of the jet head radius, $r_h(t)$, show a clear agreement with simulations (within $\sim10$--$20\%$) for both models (A and B). 
The same can be said about the time evolution of the cocoon's lateral width $r_c(t)$ 
(within $\sim 10 - 20\%$).

The time evolution of the average opening angle of the jet head $\theta_j(t)$ in simulation can be divided into two phases; with the first phase showing relatively large opening angles and unstable behavior, and the second phase showing collimated opening angles, and relatively a stable behavior.
In the first phase, the effects of the initial conditions are still present.
However, since this phase is relatively short, its contribution to the jet structure and propagation, up to the breakout, is limited.
In the second phase, which represents most of the jet propagation time, the jet head's opening angle in the analytic and semi-analytic solutions is, overall, consistent with the average opening angle in simulations (well within $\sim 50\%$). 

\subsection{The cocoon pressure $P_c$}
\label{sec:The cocoon pressure}
\begin{figure*}
    \vspace{4ex}
  \begin{subfigure}
    \centering
    \includegraphics[width=0.455\linewidth]{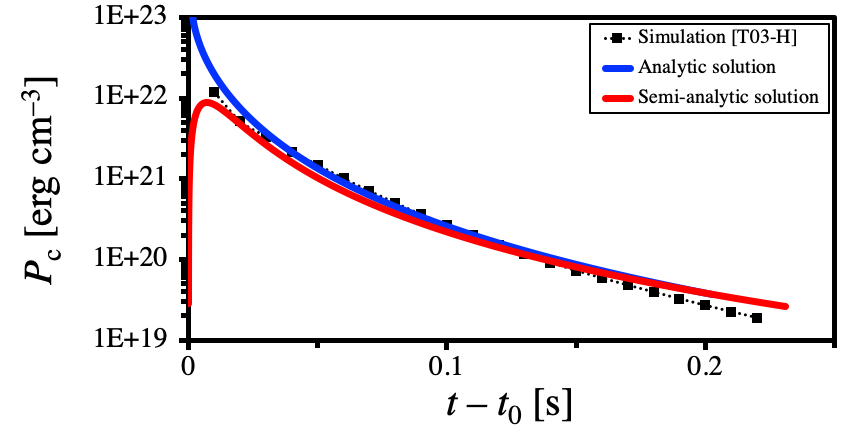} 
  \end{subfigure}
  \begin{subfigure}
    \centering
    \includegraphics[width=0.455\linewidth]{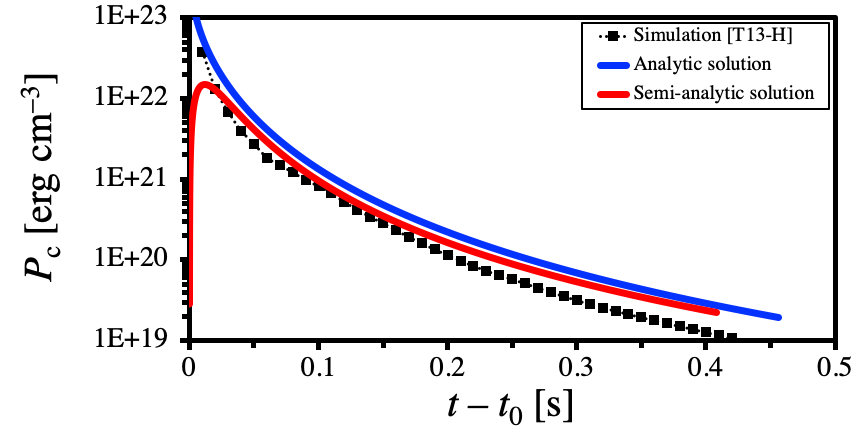} 
  \end{subfigure} 
    \begin{subfigure}
    \centering
    \includegraphics[width=0.455\linewidth]{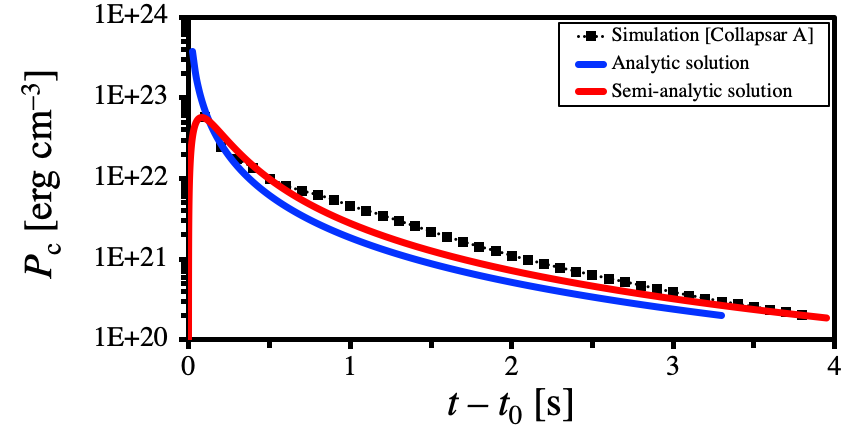} 
  \end{subfigure}
  \begin{subfigure}
    \centering
    \includegraphics[width=0.455\linewidth]{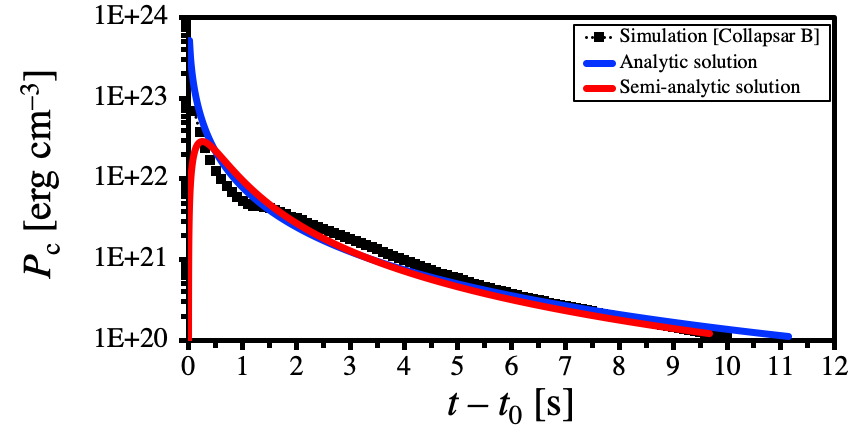}
  \end{subfigure} 
  \caption{The pressure of the cocoon $P_c$ from the jet launch time $t_0$ to the breakout time. The top two panels show two BNS merger models (T03-H and T13-H, from left to right), and the bottom two panels show two collapsar models (A and B, from left to right; see Table \ref{Table:sim}). The black dotted line with filled squares shows the average pressure in the cocoon as measured in numerical simulation. The blue line shows the cocoon pressure according to the analytic model [equation (\ref{eq:P_c approx})]. The red solid line shows the cocoon pressure according to the semi-analytic model (calculated numerically; see Section \ref{sec:Semi-analytic solution}).}
  \label{fig:P_c} 
\end{figure*}

Figure \ref{fig:P_c} shows the average cocoon pressure in simulations (measured throughout the cocoon's grid in numerical simulations and volume-averaged) compared with the cocoon pressure as inferred from our analytic and semi-analytic solutions [see Section \ref{sec:Semi-analytic solution}; and equation (\ref{eq:P_c approx}) in Section \ref{sec:Analytic solution}]. 
For both cases (BNS merger jet and collapsar jet), and for both the analytic and the semi-analytic solutions, the time evolution of the cocoon pressure up to the breakout is well consistent with numerical simulations.
This agreement indicates that the modeling presented here is a good representation of the cocoon and its interaction with the jet and the ambient medium.

Note that in the analytic solution, as the early uncollimated jet phase is not taken into account, the cocoon pressure diverges at $t\sim t_0$ due to the approximation in equation (\ref{eq:r_m>>r_m,0 app}).
The semi-analytic solution shows 
no such anomaly.

\begin{figure*}
    \vspace{4ex}
  \begin{subfigure}
    \centering
    \includegraphics[width=0.49\linewidth]{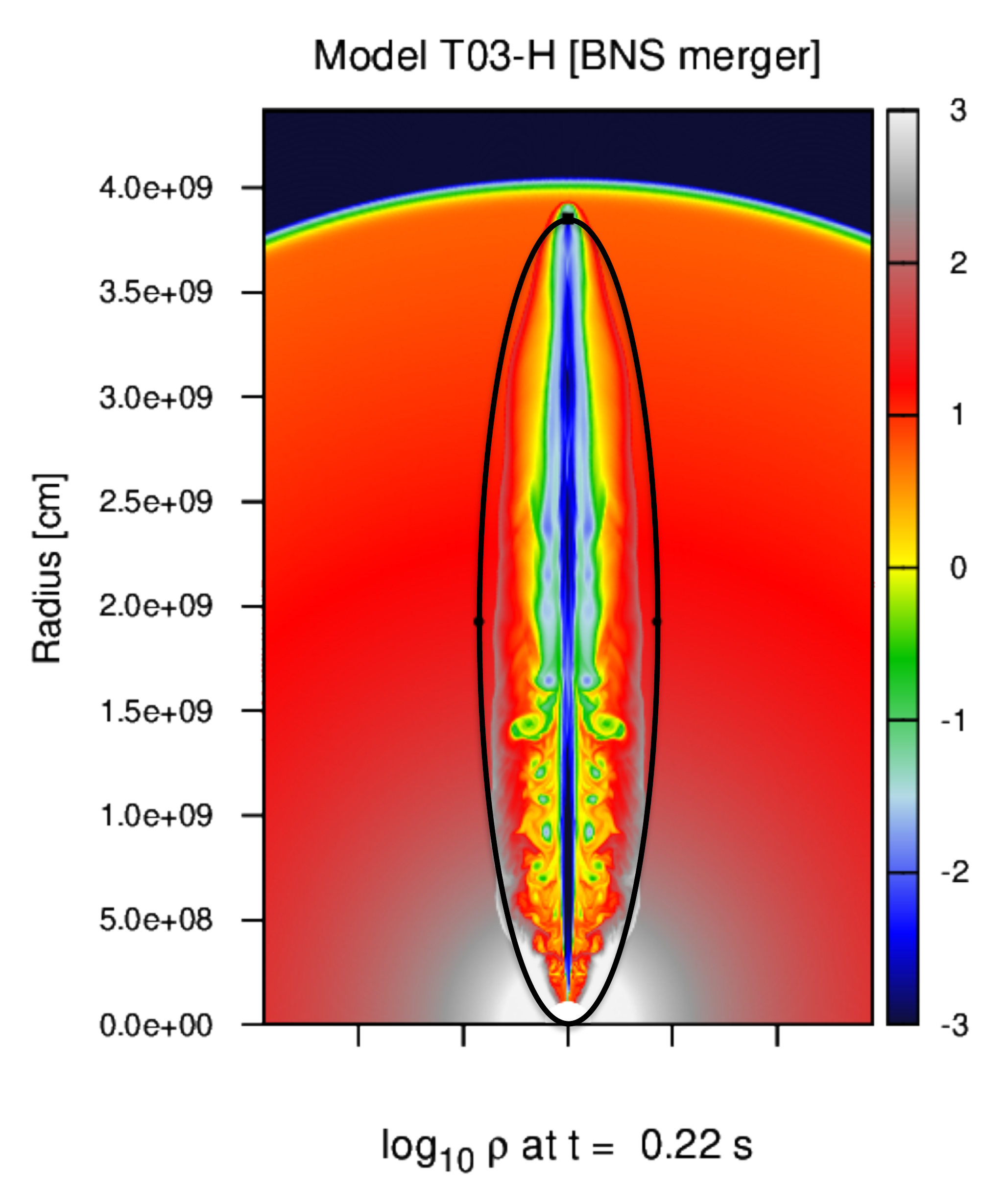} 
  \end{subfigure}
  \begin{subfigure}
    \centering
    \includegraphics[width=0.49\linewidth]{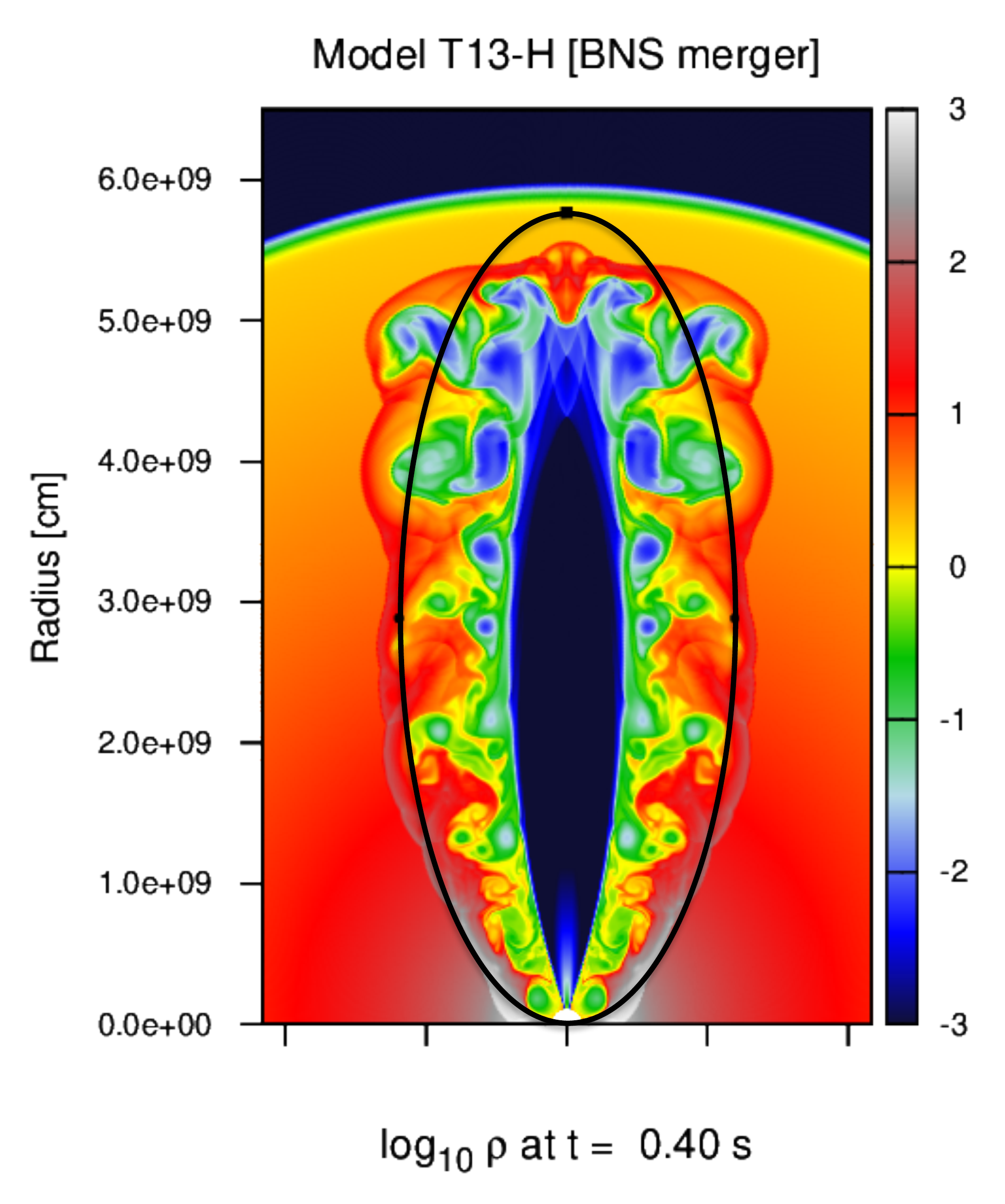} 
  \end{subfigure} 
  \vspace{4ex}\\
    \begin{subfigure}
    \centering
    \includegraphics[width=0.49\linewidth]{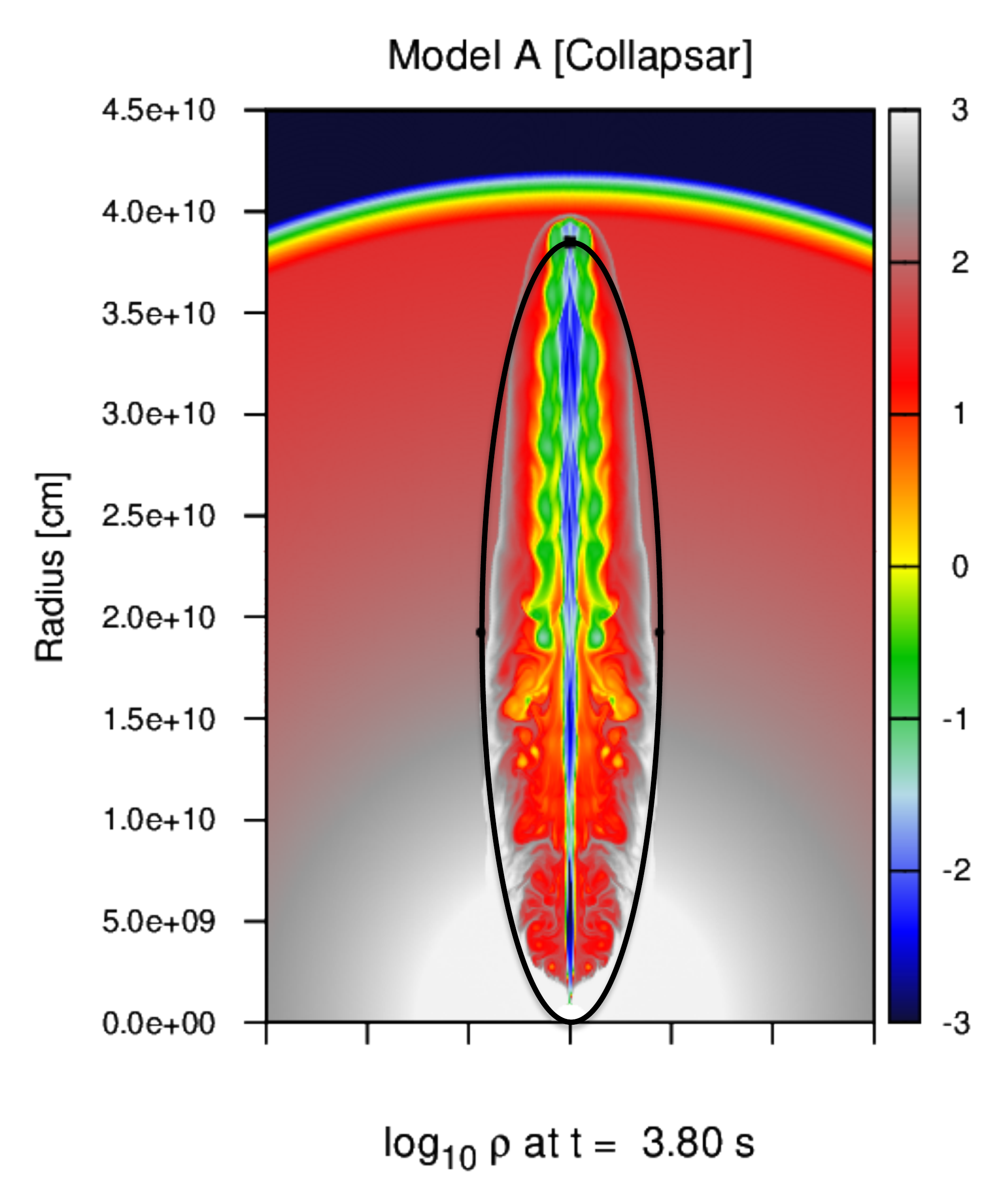} 
  \end{subfigure}
  \begin{subfigure}
    \centering
    \includegraphics[width=0.49\linewidth]{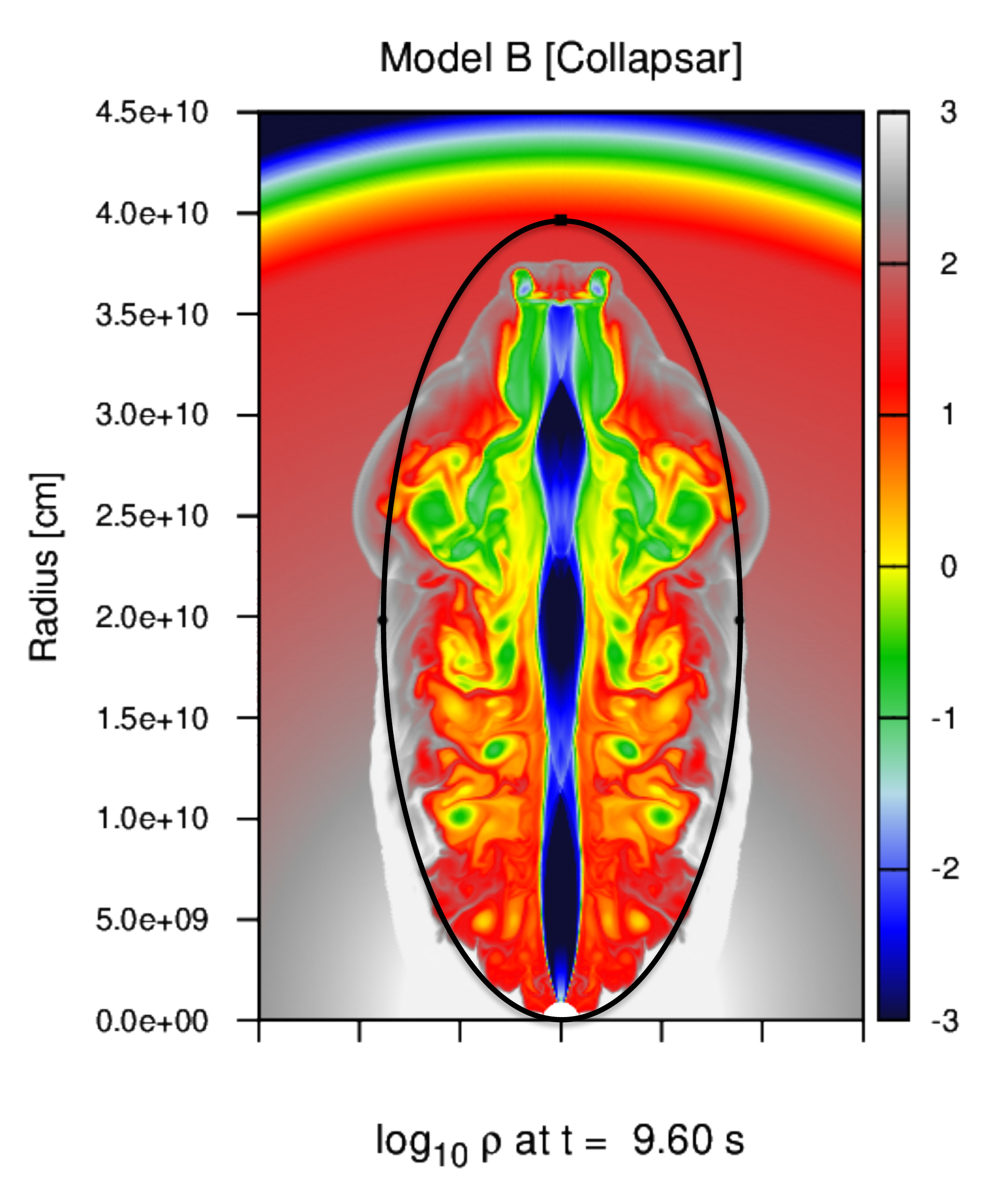} 
  \end{subfigure} 
  \caption{Density maps showing the jet-cocoon system inside the ambient medium just before the jet breakout. Four models are shown (see Table \ref{Table:sim}), where the top two models are for jet propagation in BNS merger ejecta (T03-H and T13-H), and the bottom two are for jet propagation in a stellar envelope (collapsar jets; model A and B). 
  The black filled square shows the inferred jet head radius $r_h$ by our semi-analytic model, and the two black filled circles show the inferred cocoon's lateral width $r_c$. 
  The ellipsoidal shape (solid black line) shows the jet-cocoon's shape as predicted by our modeling (the semi-analytic solution). }
  \label{fig:maps} 
\end{figure*}
\subsection{Morphology of the jet-cocoon system}
\label{sec:Morphology of the jet-cocoon system}
Figure \ref{fig:maps} presents snapshots from our numerical simulations showing the density map of the jet-cocoon system just before its breakout out of the ambient medium. 
The four models in Table \ref{Table:sim} are shown for BNS mergers (top) and collapsars (bottom).

The jet head radius and the cocoon's lateral width, 
as inferred from our semi-analytic solution, are also shown for comparison with simulations (with a black filled square, and two black filled circles, respectively).
Also, the inferred jet-cocoon morphology using the approximation of an ellipsoidal shape is shown (with a solid black line) where the semi-major axis is $\frac{1}{2}r_h$ and the semi-minor axis is $r_c$.

In Figure \ref{fig:maps}, we see a clear similarity between the morphology of the whole jet-cocoon system as inferred from our modeling and numerical simulations. 
With $r_h$ being well consistent with simulations, and the error on the cocoon's lateral width of the order of $\sim 20\%$ [at $r\sim \frac{1}{2}r_h(t)$], it can be claimed that our modeling can robustly give the cocoon volume (within $\sim 50\%$). 
Together, with the modeling of the cocoon pressure (see Section \ref{sec:The cocoon pressure}), we can conclude that all aspects of the jet-cocoon system are fairly well reproduced with our modeling.

\section{Conclusion}
\label{sec:conclusion}
In this paper we present a new jet-cocoon model.
The model is based on previous works of collapsar jet-cocoon modeling, in particular, models in \citet{2003MNRAS.345..575M}; \citet{2011ApJ...740..100B}; \citet{2013ApJ...777..162M}; \citet{2018MNRAS.477.2128H}.
From the analysis of jet propagation in numerical simulations over a wide parameter range,  
the model has been generalized to enable proper treatment of jet propagation in the case of BNS merger where the ambient medium is expanding. 
For each jet case, equations have been solved through numerical integration (semi-analytic solution), or analytically (analytic solution) after adding some approximations (see Section \ref{sec:The approximated analytic jet head velocity} and Section \ref{sec:The approximated cocoon's lateral width $r_c$}). Table \ref{Table:2.works} presents an overview of previous works on the modeling of GRB-jet propagation, and a comparison with this work. 
In summary, our results can be summarized as follows:
\begin{enumerate}
    \item Comparisons with numerical simulations show that our model's results are in a clear agreement with numerical simulations (overall, within $\sim 20\%$). 
    The time evolution of the following quantities has been shown to be well consistent with measurements from numerical simulations: the jet head radius $r_h(t)$, the cocoon radius $r_c(t)$, the jet head opening angle $\theta_j(t)$, and the cocoon pressure $P_c$ (see Figures \ref{fig:BNS case}, \ref{fig:collapsar case} and \ref{fig:P_c}).
    The cocoon's morphology, and volume, as inferred with our model are also consistent with numerical simulations (see Figure \ref{fig:maps}).
    This is the first time that results from the modeling of jet propagation in an expanding medium (as in BNS mergers) has been compared with numerical simulations over such a large set of parameters, and found consistent to such extent (see Table \ref{Table:sim}).
    \item The results of our jet-cocoon model are proven to be consistent with numerical simulations regardless of the jet case (collapsar jet or BNS merger jet), and regardless of the jet initial opening angle (see Figures \ref{fig:BNS case}, \ref{fig:collapsar case}, \ref{fig:P_c}, and \ref{fig:maps}).
    \item In addition to the semi-analytic solution, where equations are solved through numerical integration, our model offers an analytic solution, where equations are solved analytically after being approximated and simplified (see Section \ref{sec:Analytic solution}). 
    Still, we showed that, within certain conditions (e.g., $t_0-t_m \ll t_b-t_m$), the analytic solution's results are, almost, as consistent with numerical simulations as the semi-analytic model's results, despite being much simpler.
    This analytic modeling, with its simplified but fairly robust equations [e.g., equation (\ref{eq:t_b BNS}) for the breakout time], is very useful for further investigations; for instance, on the cocoon's cooling emission (Hamidani et al. in prep).
    \item The composition of the cocoon energy has been measured for both jet cases (collapsar and BNS merger), thanks to numerical simulations.
    Results show a clear contrast between the two cases;
    the cocoon energy in the case of a collapsar jet is overwhelmingly dominated by internal energy, while in the case of a BNS merger jet it is, rather, overwhelmingly dominated by kinetic energy. 
    This is the first time that such difference has been revealed. 
    As a result of this difference in internal energy, we showed that the parameter $\eta$ [see equation (\ref{eq:S3})] is much smaller ($\sim 2$ times) in the case of BNS mergers than in the case of collapsars (see Figure \ref{fig:eta}).
    Such difference has not been taken into account in previous works although it substantially affects every aspect of the jet propagation.
    Also, such difference in internal energy (in the BNS merger case) is very important when estimating the cooling emission of the cocoon; hence the importance of this result (previously mentioned in \citealt{2019ApJ...887L..16K}).
\end{enumerate}

It should be noted that the analytic modeling presented here includes several limitations.
The most important limitation is that jets here are assumed as unmagnetized. 
Other notable limitations are that effects from neutrinos, r-process, viscous wind, general relativity, stellar rotation, stellar magnetic field, etc. have been overlooked for the sake of simplicity.
Future works are likely to update our results.

Finally, it should also be noted that due to the limited computational resources, the numerical simulations presented here use the approximation of axial symmetric jets (2D) and jets are injected at relatively larger radii. 
This may result in some numerical artifacts.  
Therefore, results such as; the value of $N_s$; and the overall agreement of $\sim 20\%$ between analytic results and numerical simulations; should not be taken at face value.
We expect these values to be updated in the future once more refined numerical simulations are available. 

\begin{table*}
\caption{Comparison of jet-cocoon models in the literature. }
\label{Table:2.works}
\begin{tabular}{*6c}
\hline
     &  \multicolumn{2}{c}{Context (medium):} &  & Consistency & \\
     &  BNS mergers & Collapsars & Analytic & with  & \\
Work &  (expanding) & (static)   & solution & simulations & Comment\\
\hline
\citet{2011ApJ...740..100B} &  No    & \checkmark     & \checkmark & \checkmark & Limited to the collapsar case\\
\citet{2013ApJ...777..162M} &  No    & \checkmark     & \checkmark & \checkmark &  Limited to the collapsar case \\
\citet{2018MNRAS.475.2659M} &  No    & \checkmark$^*$ & \checkmark & ?          & $^*$Jet propagation in SLSN ejecta.\\ 
\citet{2018MNRAS.477.2128H} &  No    & \checkmark     & \checkmark & \checkmark &  Limited to the collapsar case\\
\citet{2018ApJ...866....3D} &  \checkmark  & No       & \checkmark & \checkmark & No treatment for jet collimation.\\
\citet{2018ApJ...866L..16M} &  \checkmark  & No       & \checkmark & No   & Overlooks $\eta$ and $\chi$.\\
\citet{2019ApJ...881...89L} &  \checkmark  & No       & \checkmark & ?          & Describes the jet-wind interaction.\\
\citet{2019ApJ...876..139G} &  \checkmark  & No       & \checkmark & ?          & Overlooks $\eta$ and $\chi$.\\
Salafia et al. (2020) &  \checkmark  & \checkmark     & No   & \checkmark & The effect of the expansion was not included \\
&&&&& [in $\beta_\perp$, second term in equation (\ref{eq:S2}); in $\eta$; etc.].\\
\citet{2020MNRAS.491.3192H} &  \checkmark  & \checkmark     & \checkmark & \checkmark & Simplified, after showing that $\theta_j(t)\sim \rm{Constant}$.\\
\citet{2020MNRAS.491..483L} &  \checkmark  & No       & \checkmark & No   & No treatment for jet collimation.\\
\citet{2020ApJ...895L..33B} &  \checkmark  & No       & \checkmark & No   & No treatment for jet collimation.\\
\hline
This work                   & \checkmark  & \checkmark      & \checkmark & \checkmark & \\
\hline
\end{tabular}
\end{table*}

\section*{Acknowledgements}
\addcontentsline{toc}{section}{Acknowledgements}
    We thank Amir Levinson, Atsushi Taruya, Bing Zhang, Christopher M. Irwin, Hendrik van Eerten, Hirotaka Ito, Kazumi Kashiyama, Kazuya Takahashi, Kenta Kiuchi, Kohta Murase, Koutarou Kyutoku, Masaomi Tanaka, Masaru Shibata, Ore Gottlieb, Tomoki Wada, Toshikazu Shigeyama, Tsvi Piran, and Yudai Suwa, for fruitful discussions and comments. 
    We thank the participants and the organizers of the workshops with the identification number YITP-T-19-04, YITP-W-18-11 and YITP-T-18-06, for their generous support and helpful comments. 
    
    Numerical computations were achieved thanks to the following: Cray XC50 of the Center for Computational Astrophysics at the National Astronomical Observatory of Japan, and Cray XC40 at the Yukawa Institute Computer Facility.
   
    This work is partly supported by JSPS KAKENHI nos. 20H01901, 20H01904, 20H00158, 18H01213, 18H01215, 17H06357, 17H06362, 17H06131 (KI). 

\section{Data availability}
The data underlying this article will be shared on reasonable request to the corresponding author.




\bibliographystyle{mnras}
\bibliography{04c-mnras} 


\appendix
\section{Energy composition of the cocoon}
\label{sec:A}
The total energy of the cocoon can be written as:
\begin{equation}
E_{c} = E_i + E_{k,in} + E_{k,0}.
\label{eq:E/Ec}
\end{equation}
$E_i$ is the internal energy remaining in the cocoon, out of the energy injected by the central engine into the cocoon (through the jet head).
$E_{k,in}$ is the part of the kinetic energy of the cocoon that originates from the energy injected by the engine into the cocoon (even if the injected energy by the engine is in the form of internal energy, a part of it gets converted into kinetic energy; inside the shocked jet, before reaching the cocoon; or later, inside the cocoon, due to the adiabatic expansion of the cocoon). 
These two energies ($E_i$ and $E_{k,in}$) summed up equal the total energy delivered by the engine into the cocoon: 
\begin{equation}
E_i+E_{k,in}=(t-t_0)L_j(1-\langle{\beta_h}\rangle) .
\label{eq:E_in}
\end{equation}
Finally, $E_{k,0}$ is the kinetic energy initially carried by the part of the ambient medium that became the cocoon.
It can be calculated with following integration:
\begin{equation}
E_{k,0}= \int_{cocoon} (\Gamma_a -1)\rho_a dV.
\label{eq:E_k0}
\end{equation}
Note that, $E_{k,0}=0$ in the collapsar case as the medium is static. 

\begin{figure*}
    \vspace{4ex}
  \begin{subfigure}
    \centering
    \includegraphics[width=0.495\linewidth]{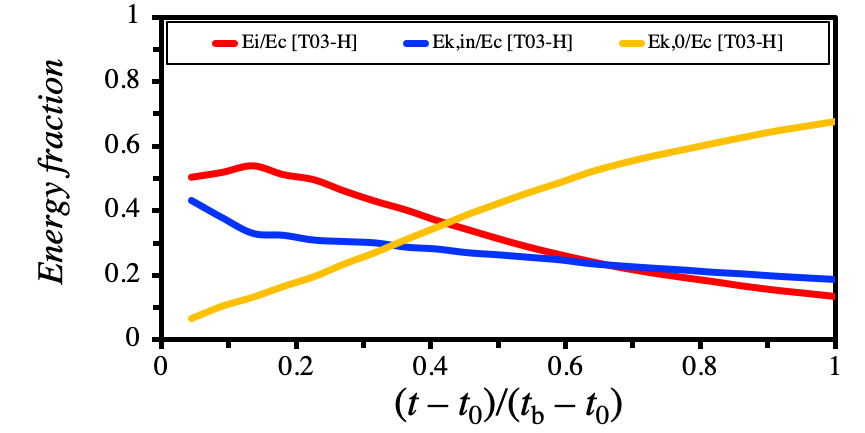}
  \end{subfigure}
  \begin{subfigure}
    \centering
    \includegraphics[width=0.495\linewidth]{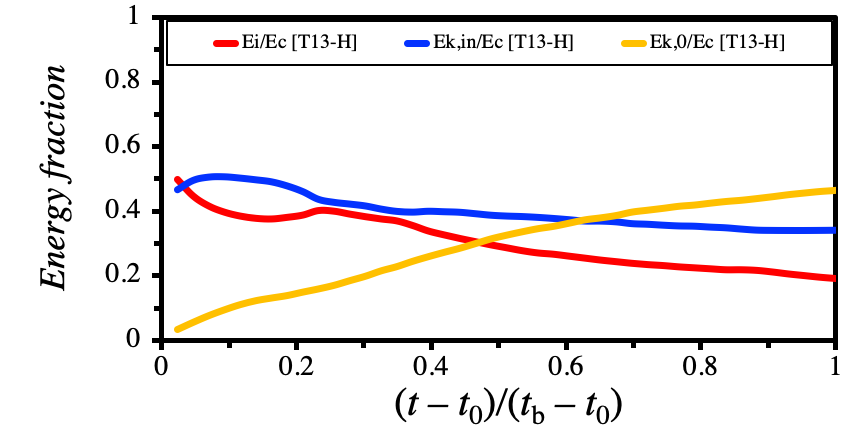} 
  \end{subfigure} 
    \begin{subfigure}
    \centering
    \includegraphics[width=0.495\linewidth]{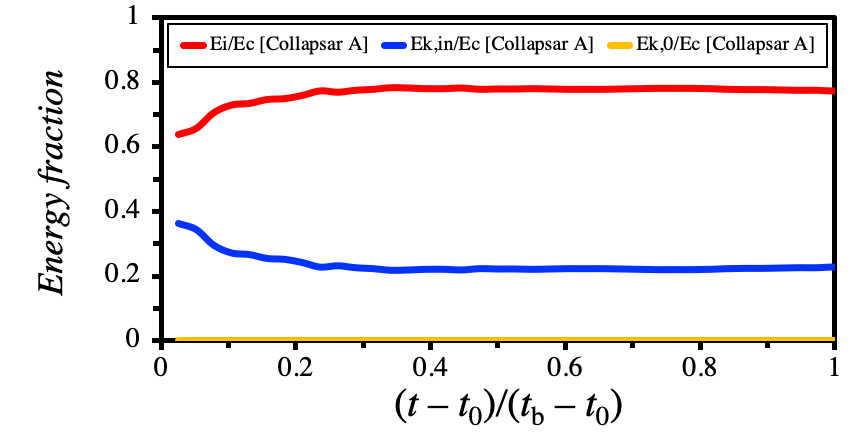} 
  \end{subfigure}
  \begin{subfigure}
    \centering
    \includegraphics[width=0.495\linewidth]{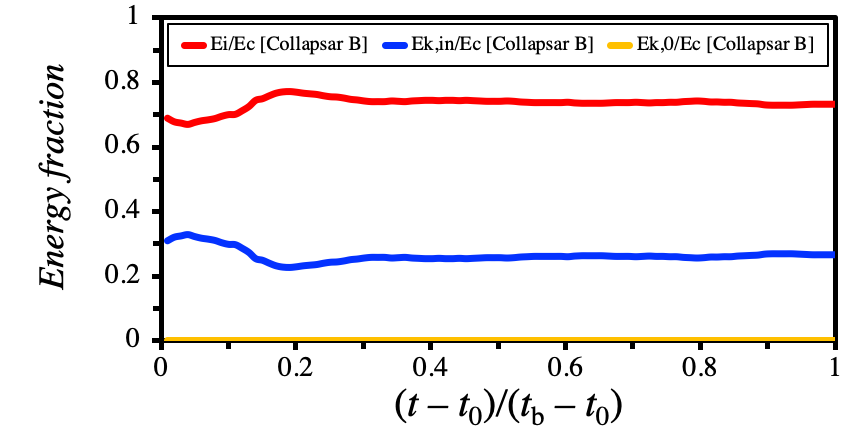} 
  \end{subfigure}
  \caption{Energy composition of the cocoon, from the jet launch time $t=t_0$, to the jet breakout time $t=t_b$, as measured from numerical simulations. 
  $E_i/E_c$ is the fraction of the internal energy remaining in the cocoon (same as the previously presented in the top panel of Figure \ref{fig:eta}), 
  $E_{k,in}/E_c$ is the fraction of the kinetic energy originating from the central engine in the cocoon (mostly injected as internal energy, but later converted into kinetic energy), 
  and $E_{k,0}/E_c$ is the fraction of the kinetic energy that was initially carried by the part of the ambient medium that became the cocoon (unlike $E_{k,in}$, it did not originate from the central engine) in the cocoon [see equation (\ref{eq:E/Ec}) and the explanation that follows]. 
  The top two panels are for BNS merger jet models (from left to right, T03-H and T13-H), and the bottom two panels are for collapsar jet models (from left to right, A and B). 
  See Table \ref{Table:sim} for more details about each of the four models.}
  \label{fig:energy} 
\end{figure*}

The fraction of the internal energy in the cocoon has been presented in Figure \ref{fig:eta}.
Figure \ref{fig:energy} offers more details by showing the fraction of each of the three energies [in equation (\ref{eq:E/Ec})], as measured from numerical simulations, and for each of the four models (see Table \ref{Table:sim} for the parameters of each model). 

As it can be seen from Figure \ref{fig:energy}, numerical simulations show that, in the case of BNS merger jets, the energy contribution of the initial kinetic energy of the ambient medium ($E_{k,0}$) in the cocoon energy increases with time and is substantial at later times ($t\sim t_b$).
This is because of the increasing volume of the cocoon with time (since, $E_{k,0}\propto V_c$).
In the collapsar jet case, the situation is simpler as $E_{k,0}=0$.

In the case of BNS merger jets, the energy contribution of both the internal energy ($E_{i}$) and the kinetic energy that originates from the central engine ($E_{k,in}$) declines with time (while it is constant over time in the collapsar jet case).
This is due to two main reasons. 
First, the increasing contribution of $E_{k,0}$ in $E_c$ with time, which ends up lowering the share of $E_{i}$ and $E_{k,in}$ in $E_c$. 
Second, the higher jet head velocity (in the central engine frame) in the BNS merger jet case (relative to the collapsar jet case) [see equations (\ref{eq:v_h BNS}) and (\ref{eq:v_h collapsar})]; 
this implies that in the BNS merger jet case, and relative to the collapsar jet case, the fraction of the energy injected by the central engine that ends up in the cocoon is smaller, and the energy that is contained inside the jet and never reaches the cocoon is higher [$\langle{\beta_h}\rangle$ is higher in the BNS merger case \citealt{2018PTEP.2018d3E02I} $\sim {2 v_m}/{c}$; see equation (\ref{eq:E_in})].

Another notable difference between the BNS merger jet case and the collapsar jet case is the ratio $E_i/E_{k,in}$.
In the BNS merger jet case, $E_i$ is comparable to $E_{k,in}$ (and even slightly smaller at $t\sim t_b$) while in the collapsar jet case $E_i$ is much higher than $E_{k,in}$ ($\sim 80\%$ and $\sim 20\%$, respectively).
This can be explained by the continues expansion of the medium in the BNS merger jet case, resulting in the conversion of a substantial part of the internal energy $E_i$ into kinetic energy (into $E_{k,0}$). 
Roughly, it is estimated that about half of the cocoon's internal energy is lost with this process.

To conclude, we showed, for the first time (to our best knowledge), that the fraction of internal energy in the cocoon, $E_i/E_c$, is much smaller in the case of BNS merger jet (relative to the case of collapsar jet) due to following three reasons: 
i) the increasing contribution of the initial kinetic energy of the ambient medium, $E_{k,0}/E_c$, in the cocoon of a BNS merger jet case; 
ii) the high jet head velocity (in the frame of the central engine) in the BNS merger jet case; 
and iii) the adiabatic expansion process in the cocoon of a BNS merger jet case (due to continues expansion of the ambient medium).

The fraction $E_i/E_c$ is an important quantity as it strongly affects the jet propagation by affecting the value of $\eta$ (and $\eta'$) [\citealt{2011ApJ...740..100B}], hence the importance of these results.


\section{Energy composition of the cocoon: testing BNS models with a hypothetical static ambient medium}
Here, our explanation of the origin of the difference in the energy composition of the cocoon, between the collapsar case and the BNS merger case (being different due to the adiabatic expansion of the cocoon, and due to the contribution from the initial kinetic energy of the ambient medium, in the BNS merger case, hence being due to the expanding aspect of the ambient medium; see Section \ref{sec:Mesurements of internal energy}; and Appendix \ref{sec:A}) is tested.
We present two test models T03-H-static, and T13-H-static.
These two models share the same parameters as the models T03-H, and T13-H (respectively), except that the ambient medium velocity is set as $v_m=0$ (hence the naming ``-static'').
This test is carried out to make sure that the overwhelming difference between the BNS merger case and the collapsar case, in terms of the cocoon's energy composition, is not caused by other parameters (such as the parameters of the jet or the ambient medium; e.g., $L_j$, or $M_a$). 

\begin{figure}
    \vspace{4ex}
    \includegraphics[width=0.995\linewidth]{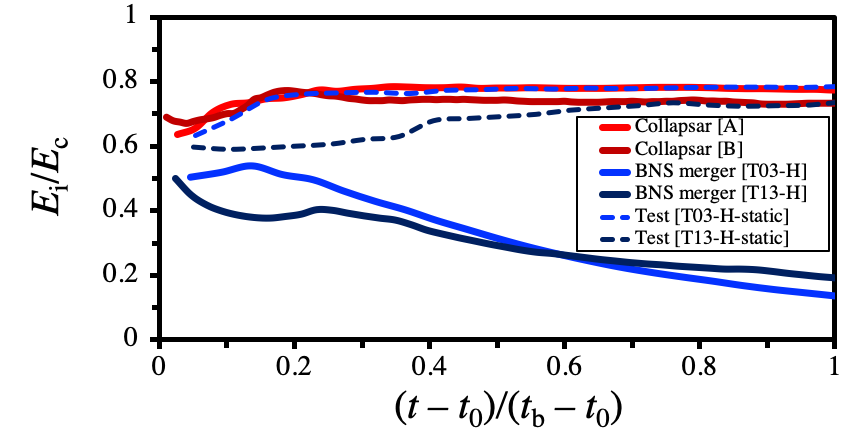} 
  \caption{Same as the top panel in Figure \ref{fig:eta}, with the addition of two test models (T03-H-static and T13-H-static) having the same parameters as the BNS merger case models (T03-H and T13-H, respectively) expect that their ambient medium is set as static ($v_m=0$, while in T03-H and T13-H $v_m=0.345c$; see Table \ref{Table:sim}).
  }
  \label{fig:test} 
\end{figure}

In Figure \ref{fig:test}, the two models T03-H-static and T13-H-static (with $v_m=0$), compared with the two models T03-H and T13-H ($v_m=0.345c$), show that the expanding aspect of the ambient medium is, indeed, the origin of the large difference in the cocoon's energy composition (mostly internal for a static medium and mostly kinetic for an expanding medium).
This is in perfect agreement with our explanation in Section \ref{sec:Mesurements of internal energy} and in Appendix \ref{sec:A}.

\section{The calibration coefficient $N_s$}
\label{sec:C}
The calibration coefficient, $N_s$, was first introduced by \citet{2018MNRAS.477.2128H} after the realization (first made by \citealt{2013ApJ...777..162M}) that the analytic model by \citet{2011ApJ...740..100B} needs to be corrected to match the simulation results.
This parameter is used to correct the analytic (or semi-analytic) value of $\tilde{L}$ (to $\tilde{L}_c$), in accordance with numerical simulations, as follows:
\begin{equation}
    \tilde{L}_c=N_s^2 \tilde{L}.
\end{equation}
$N_s$ was studied by \citet{2018MNRAS.477.2128H} for the case of collapsar jets.
By measuring $\tilde{L}$ in numerical simulations, it was found that $N_s$ should take values in the range $\sim \frac{1}{2.5} - \frac{1}{3}$ in the non-relativistic (Newtonian) regime, and $\sim 1$ in the relativistic regime (see Figure 12 in \citealt{2018MNRAS.477.2128H}). 

Here, the value of $N_s$ is deduced based on the jet breakout time;
the value of $N_s$ calibrates $\tilde{L}$ (to $\tilde{L}_c$) so that the semi-analytic (or analytic) breakout time matches the numerical simulation's breakout time.
Hence, with this procedure, the values of $N_s$ found here can be understood as average values --
throughout the jet propagation from the jet launch time ($t_0$) to the jet breakout time ($t_b$).

In this study, the jet propagation is solved for the case of BNS mergers where the ambient medium is expanding, in addition to the case of collapsars.
We adopt the analytic model of \citet{2011ApJ...740..100B}, with a few notable modifications.
In particular, the modeling of the cocoon's lateral width ($r_c$ and $\beta_\perp$), the cocoon morphology and volume ($V_c$), and the fraction of internal energy in the cocoon $E_i/E_c$ (or $\eta'$) have been modified [see equations (\ref{eq:V_c}); (\ref{eq:S2}); and Table \ref{Table:sim}; respectively]. 
These modifications are expected to affect the value of $N_s$.

The system of equations of the jet propagation is solved in two methods:
\begin{itemize}
    \item Semi-analytic solution: the equations are solved through numerical integration (see Section \ref{sec:Semi-analytic solution}). 
    \item Analytical solution: approximations are added to the semi-analytic solution to simplify it to a fully analytic form (see Section \ref{sec:Analytic solution}).
\end{itemize}
Due to the approximations introduced in the analytic solution, $N_s$ is not exactly the same in these two methods (see Appendix \ref{sec:C2}).

\subsection{$N_s$ in the semi-analytic solution}
\label{sec:C1}
From the semi-analytic modeling (see Section \ref{sec:Semi-analytic solution}), we get $N_s=0.53$ for the collapsar case ($\tilde{L} \sim 0.08 - 1$), and $N_s=0.75$ for the BNS merger case ($\tilde{L} \sim 0.2 - 0.7$).

First, the value of $N_s$ in the collapsar case is slightly different from the value suggested by \citet{2018MNRAS.477.2128H} ($N_s\approx 0.33 - 0.40$ in the non-relativistic case; see their Figure 12 in the range $\tilde{L}_a\lesssim1$).
This can be explained by differences in the modeling [of the cocoon's lateral width ($\beta_\perp$ and $r_c$), the cocoon's volume ($V_c$), the fraction of internal energy in the in the cocoon ($E_i/E_c$), and $\eta$ (or $\eta'$)].

Second, $N_s$ in the BNS merger case is substantially different than in the collapsar case.
This difference is mainly related to the expanding aspect of the ambient medium, which is found to strongly affect parameters such as $E_i/E_c$ (or $\eta'$; see Figure \ref{fig:eta}), and hence affecting the cocoon pressure on the jet.

\subsection{$N_s$ in the analytic solution}
\label{sec:C2}
Using the analytic solution (see Section \ref{sec:Analytic solution}) we get $N_s=0.38$ for the collapsar case, and $N_s=0.46$ for the BNS merger case.

In the collapsar case, the analytic solution is overall very similar to the semi-analytic solution.
One major difference is the absorption of the term $(1 + \tilde{L}_c^{1/2})^{-1} \lesssim 1$ into $N_s$ [see Section \ref{sec:The approximated analytic jet head velocity} and equation (\ref{eq:beta_h 2 approx collapsar})].
This is why $N_s$ here differs from $N_s$ in the semi-analytic solution.
$N_s$ can be estimated for our analytic solution based on our results in the semi-analytic solution ($\tilde{L} \sim 0.08 - 1$, and $N_s = 0.53$, for models A and B; see Appendix \ref{sec:C1}) as follows:
\begin{equation}
    N_s \approx 0.53 \:(1+0.53 \: \tilde{L}^{1/2})^{-1}.
\label{eq:N_s analytic collaspar}
\end{equation}
Using these values in the above equation, it can be confirmed that, overall, $N_s \sim 0.3 - 0.4$.

In the BNS merger case the difference in $N_s$ is more substantial;
$N_s=0.75$ for the semi-analytic solution, and $N_s = 0.46$ for the analytic solution.
This is because, in addition to the term $(1 + \tilde{L}_c^{1/2})^{-1}$, 
the term $(1-\beta_a)$ [usually $\sim 0.6 - 0.9$] and the term $\frac{1}{\Gamma_a}$ have also been absorbed into $N_s$ [see Section \ref{sec:The approximated analytic jet head velocity} and equation (\ref{eq:beta_h 2 approx}); also see equations (\ref{eq:L expression approx}) and (\ref{eq:N_s})].
Hence, $N_s$ for the analytic solution (BNS merger case) can be estimated using the results of the semi-analytic solution $N_s$, $\beta_a$, and $\tilde{L}$, as follows:
\begin{equation}
    N_s \approx 0.75 \left[ \frac{1-\beta_a}{(1+0.75\:\tilde{L}^{1/2})(1-\beta_a^2)^{1/2}}\right],
\label{eq:N_s analytic BNS}
\end{equation}
where $\tilde{L} \sim 0.2 - 0.7$ and $\beta_a \sim 0.2$ gives roughly $N_s \sim 0.4$ (for T03-H and T13-H; see Appendix \ref{sec:C1}) \footnote{It should be noted that more difference is caused by the approximation introduced in equation (\ref{eq:r_m>>r_m,0 app}); 
although, this is largely (but not entirely) mitigated by focusing on cases where $t_b-t_m\gg t_0-t_m$.}.

\subsection{$N_s$ dependence on the parameter space}

It should be noted that $N_s$ can take different values beyond the parameter space studied here (i.e., $\tilde{L} \sim 10^{-1} - 1$; and $v_m = 0.35c$).
Therefore, the values of $N_s$ given here should not be taken at face value; 
$N_s$ should always be associated with its parameter space, 
and calibrated with simulations (if possible) if used outside of this parameter space. 

Nevertheless, we believe that at the close vicinity of our parameter space, and when using the semi-analytic (or analytic) model presented here, 
the values of $N_s$ given here should be reliable.

\bsp	
\label{lastpage}
\end{document}